\documentclass[fleqn,10pt]{wlscirep}
\usepackage[utf8]{inputenc}
\usepackage[T1]{fontenc}
\usepackage{subcaption}
\usepackage{graphicx}
\title{Wave energy converters as offshore wind farm guardians: a pathway to resilient ocean systems}

\author[1,*]{Olivia Vitale}
\author[2]{Maha N. Haji}
\affil[1]{Cornell University, Sibley School of Mechanical and Aerospace Engineering, Ithaca, NY, 14850, United States}
\affil[2]{University of Michigan, Department of Mechanical Engineering, Ann Arbor, MI, 48109, United States}

\affil[*]{ov35@cornell.edu}


\begin{abstract}
Maximizing the durability and reliability of offshore wind farms is essential for the clean energy transition. In this work, we demonstrate how wave energy converter (WEC) farms can shelter offshore wind farms from cyclic wave loading, resulting in significant reductions in turbine fatigue damage. Using experimentally validated hydrodynamic models, we show that WEC arrays can reduce wave-induced fatigue damage on the turbines by up to 25\%, potentially lowering required monopile diameters and extending turbine lifetimes. This damage reduction propagates to the levelized cost of energy (LCOE) of the wind farm, targeting cost reductions in nearly 50\% of the total system costs. Additionally, WEC farms can benefit from this co-location by sharing siting costs, operation and maintenance teams, and mooring and transmission cables with the offshore wind farm. This work advances the design of integrated wind–wave systems, supporting resilient, cost-effective offshore renewables for global deployment.

\end{abstract}
\begin{document}

\flushbottom
\maketitle
%
%
\thispagestyle{empty}

\section*{Introduction}
The power production potential for offshore wind has led to large current and projected growth in global installed capacity\cite{DIAZ2020107381,GWECreport}. Despite its relevance to the modern grid, the intermittency of wind power results power quality disturbance to the grid, harming electrical components\cite{SHAO2024114094}. Additionally, installing wind turbines offshore introduces significant additional fatigue stresses due to wave loading, reducing the turbine's lifetime\cite{SUN2019472,SAENZAGUIRRE2022116303}. For fixed-bottom offshore wind, wave-induced fatigue loads have caused monopile diameters to increase around 60\%\cite{ACHMUS2009725} or required developers to use jacket structures \cite{HAFELE201899}, increasing capital costs. These problems are exacerbated for floating offshore wind, where ocean waves also induce loads on mooring cables and floaters and cause oscillations of the turbines themselves\cite{PUSTINA2020109984,en6084097}. Deploying a wave energy converter (WEC) farm upstream of an offshore wind farm can dampen incoming waves before they reach the downstream turbines. This shelters the turbines from cyclic wave loading while providing power smoothing to the grid.

The intermittency of the wind resource increases the required storage capacity and reduces power reliability, pointing to a need for power smoothing \cite{REN201747}. Several studies have proposed co-locating WEC and wind farms for power extraction symbioses\cite{VEIGAS2014300,DELPOZOGONZALEZ2023127176,PEREZCOLLAZO2015141,MICHAILIDES2016675,jariwla,zhoanat,sebastian}. Kluger et al.\cite{KLUGER2023120389} studied a combined WEC and offshore wind farm with 100 6~MW turbines and a variable number of 286~kW WECs. They found a 15\% smoother power output when 50\% of the turbines had a WEC attached to their platform compared to a standalone wind farm. Additionally, Astariz et al. \cite{ASTARIZ201563} investigated the Alpha Ventus wind farm, composed of 12 5~MW turbines. They found an increase in allowable windows for maintenance due to the presence of an upstream WEC array containing between nine and 32 WaveCat\cite{FERNANDEZ201258} devices. This further motivates the co-location of these offshore systems, as WECs provide additional benefits beyond wave protection.

Many studies have evaluated WEC sheltering for coastlines or offshore aquaculture farms. Flanagan et al. \cite{Flanagan2022} studied the effect of a WEC array on coastal ocean environment and sediment transport, finding wave height reduction to be a function of distance between WECs and the number of WEC devices. They found closer devices (one diameter apart) resulted in 30\% greater spatial impact and more devices (10-15) resulted in 200-300~m longer shore coverage. Chang et al. \cite{CHANG2016636} investigated several WEC architectures to determine the coastal sheltering effect of each. They found a maximum of 30\% wave height reduction using a WEC array with 100 bodies, with other cases resulting in less than a 15\% decrease in wave height. Venugopal et al. \cite{VENUGOPAL201737} found large wave height reductions immediately in lee of the WEC array, but a lessening effect as the distance from the farm increased.

Though these studies provided a baseline for co-located wind-wave farms, their modeling methods face serious limitations and inaccuracies due to their simplifying assumptions. Current wave sheltering studies simplify the WEC's dissipative properties, assuming the only energy removed from the waves is energy the device captures. Contardo et al. \cite{CONTARDO2018175} used this power capture method to model their WEC's sheltering capabilities. They then completed two deployments of their physical WEC device, concluding this method was poor at predicting actual wave sheltering. This is due to the method ignoring the energy that WECs scatter and dissipate through fluid-structure interactions, separate from the energy they capture. A few studies have attempted wave sheltering characterization methods that account for the fluid-structure interactions \cite{ATAN2019373, Silva2018, CARBALLO2013216}. However, they do not include interactions between WEC devices and are only conducted in nearshore regions, which represent different fluid physics than locations hosting offshore wind farms. A few models have investigated WEC sheltering explicitly for offshore wind \cite{10.1115/IOWTC2018-1077,clark-2}, but only employ the power capture method.

This study analyzes the WEC array wave sheltering performance based on fluid-structure interactions rather than absorbed power. With our experimentally validated method, the full picture of energy propagation through the WEC farm is captured, providing an accurate estimate of wave sheltering. We measure wave height up and downstream of the devices to quantify the dissipative coefficients. This is standard practice in coastal engineering \cite{Seelig1980, Bao2022} but has been neglected in WEC sheltering studies. 

Additionally, all WEC sheltering studies to date have neglected hydrodynamic interactions between WEC devices in a farm. WECs disturb the nearby wave field and, when configured in arrays, disturb the motion of the nearby devices \cite{BABARIT201368, balitsky}. These interactions are relevant for determining device motion and power extraction performance\cite{BABARIT201368}; however, the impact of these interactions on the larger ocean environment has not been studied. Zou et al. \cite{ZOU2024120719} attempted to capture WEC array wake effects by accounting for power extracted throughout the array, but still neglected hydrodynamic interactions. This work shows the relevance of including these interactions in the model, as they significantly effect device behavior across a range of sea states.

Existing studies assume power-capture–only dissipation and neglect scattering and multi-body interactions. We show these assumptions bias architectural comparisons and severely underpredict sheltering performance. This work numerically models two different WEC architectures in an optimized array configuration. The hydrodynamic interactions were experimentally validated, and empirical corrections were included in the model. The WEC devices presented in this study were physically built at the 1:50 Froude scale and tested in a deep water, non-breaking, non-Stokes fluid regime. The validated model combines a boundary element method (BEM) solver for the near-field hydrodynamic interactions and energy propagation coefficients with a spectral action balance model for far-field wave height reduction. The presented model:
\begin{enumerate}
    \item can evaluate any general WEC architecture unbiased by rated power,
    \item accounts for hydrodynamic interactions, and 
    \item defines the relationship between wave height reduction and fatigue damage reduction on offshore wind turbines.
\end{enumerate}
\begin{figure}[ht!]
    \begin{subfigure}[b]{0.45\textwidth}
        \centering
        \includegraphics[width=0.90\linewidth]{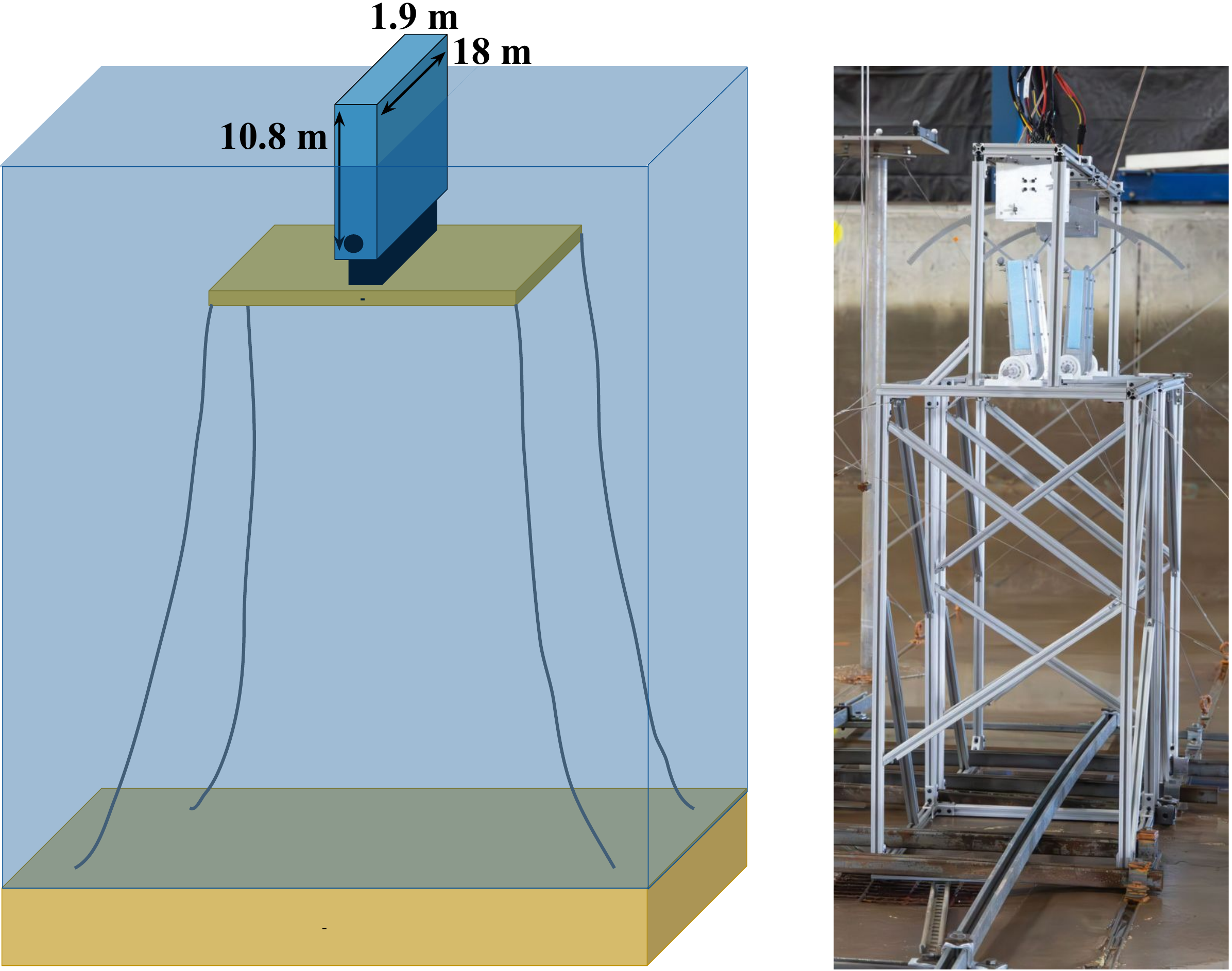}
        \caption{OSWEC}
    \end{subfigure}
    \hfill
    \begin{subfigure}[b]{0.45\textwidth}
        \centering
        \includegraphics[width=0.85\linewidth]{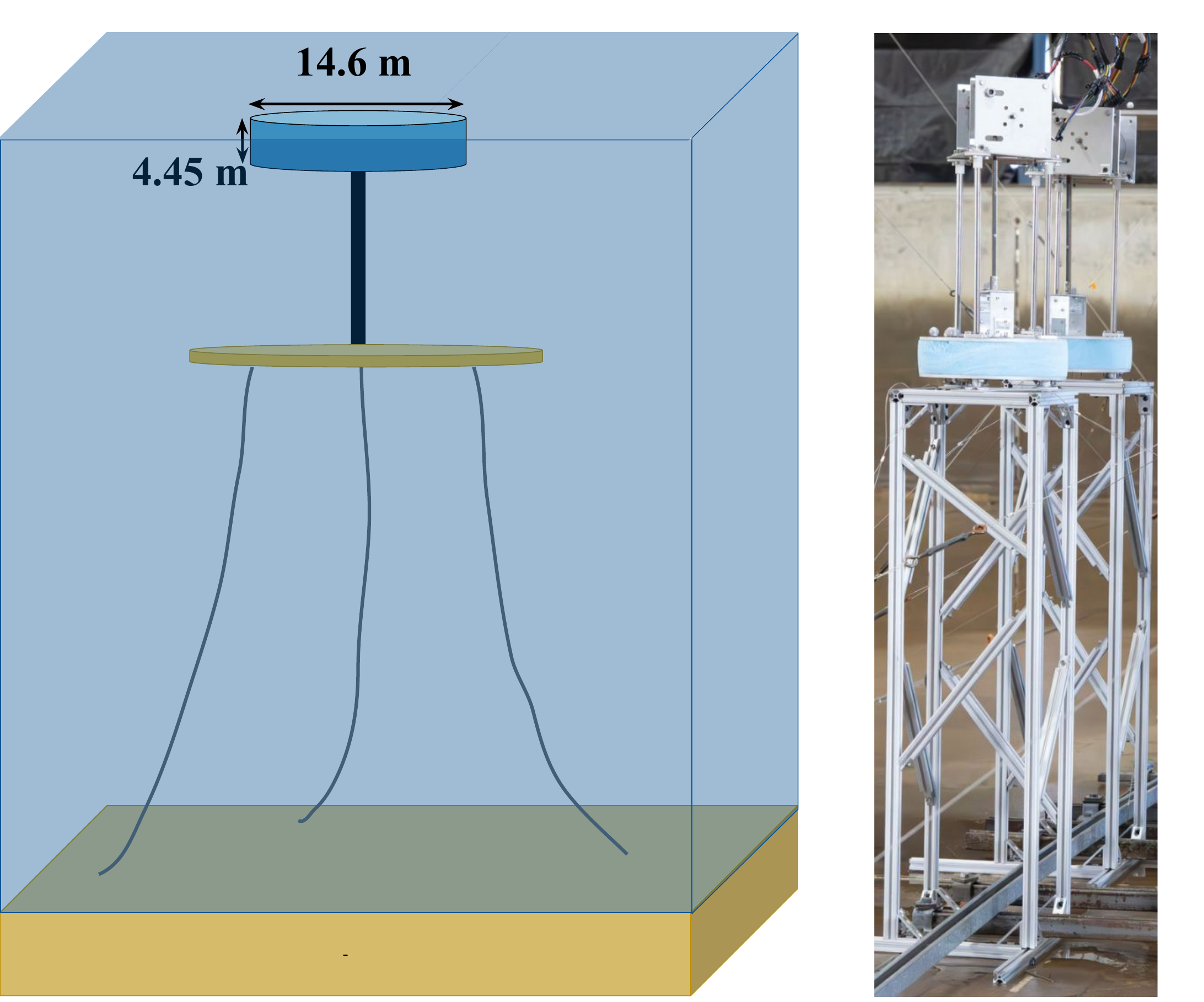}
        \caption{PA}
    \end{subfigure}
    \caption{Depiction of the two common WEC architectures evaluated in this study. Relevant dimension are provided in the drawing, and images of the physical prototypes are shown next to the drawing for context.}
    \label{fig:archetypes}
\end{figure}

\subsection*{WEC Devices}
The two WEC architectures evaluated in this study are shown at full ocean-scale in Fig. \ref{fig:archetypes}. The first architecture is called an oscillating surge WEC (OSWEC), a flap which extracts power from the waves through oscillatory pitching motion. The second architecture is a heaving point absorber (PA), a cylindrical buoy extracting power from the ocean waves by following their up-and-down (heave) motion. The OSWEC's width orthogonal to the incident wave direction is 18~m, while the PA's is 14.6~m. These dimensions were chosen based on the physical, tank-scale prototypes, built at the 1:50 Froude scale. Details on the design and build of these prototypes can be found in Vitale et al.\cite{vitalebuild} The natural oscillation period of the OSWEC is 21~s, while the PA's is 5.09~s. Both architectures employed a mechanical rack-and-pinion power take-off (PTO) mechanism to drive a motor-generator, shown in the Appendix. The PTO is how the devices extract energy from the waves, proportional to the damping present in the PTO. Each device's mechanical power extraction at full ocean-scale was estimated based on the tank-scale experiments. Each WEC's power extraction was different based on its position in the array. This is due to the hydrodynamic interactions between devices, where some devices experience motion amplification and others see decreased performance. The OSWECs predicted power extraction at the peak spectral period (11~s) is between 145 and 932~W based on position in the array. For the PAs, the mechanical power extraction is between 86.2 and 523~W.

\section*{Results}
\subsection*{Wave Sheltering Performance}
\begin{figure}[ht!]
     \begin{subfigure}[b]{0.49\textwidth}
         \centering
         \includegraphics[width=\linewidth]{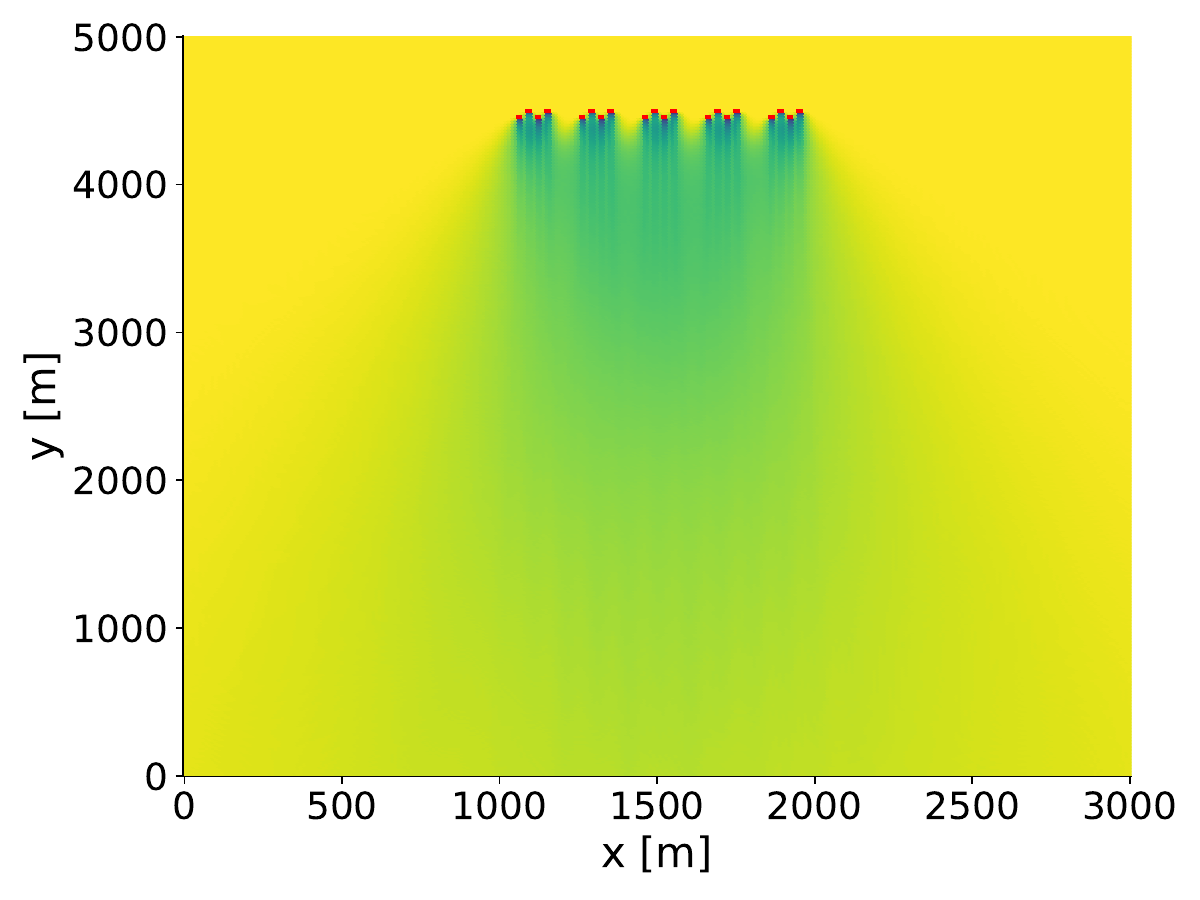}
         \caption{20-body OSWEC array}
         \label{fig:osSWANcont}
     \end{subfigure}
     \hfill
     \begin{subfigure}[b]{0.55\textwidth}
         \centering
         \includegraphics[width=\linewidth]{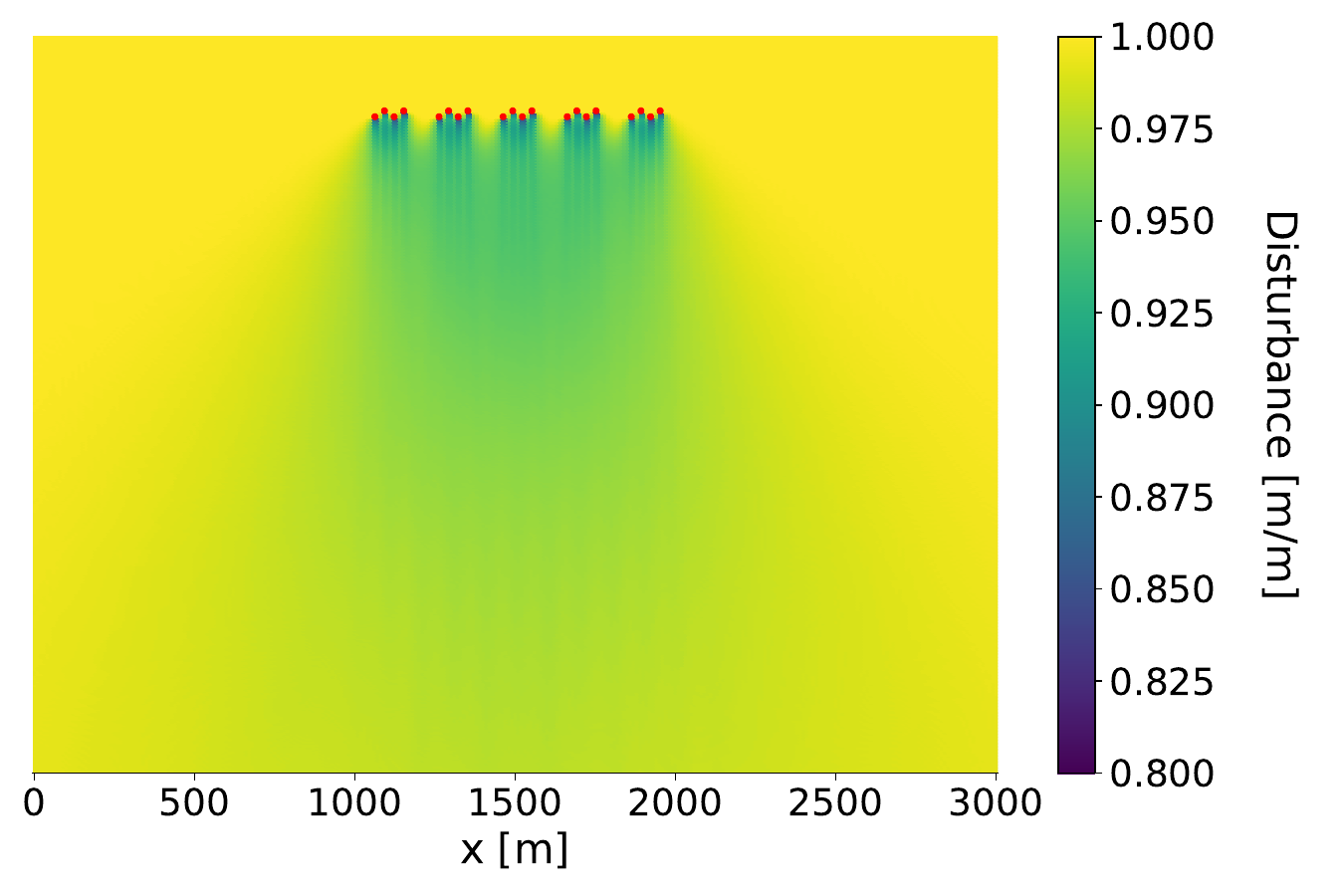}
         \caption{20-body PA array}
         \label{fig:paSWANcont}
     \end{subfigure}
     \hfill
     \begin{subfigure}[b]{0.49\textwidth}
         \centering
         \includegraphics[width=\linewidth]{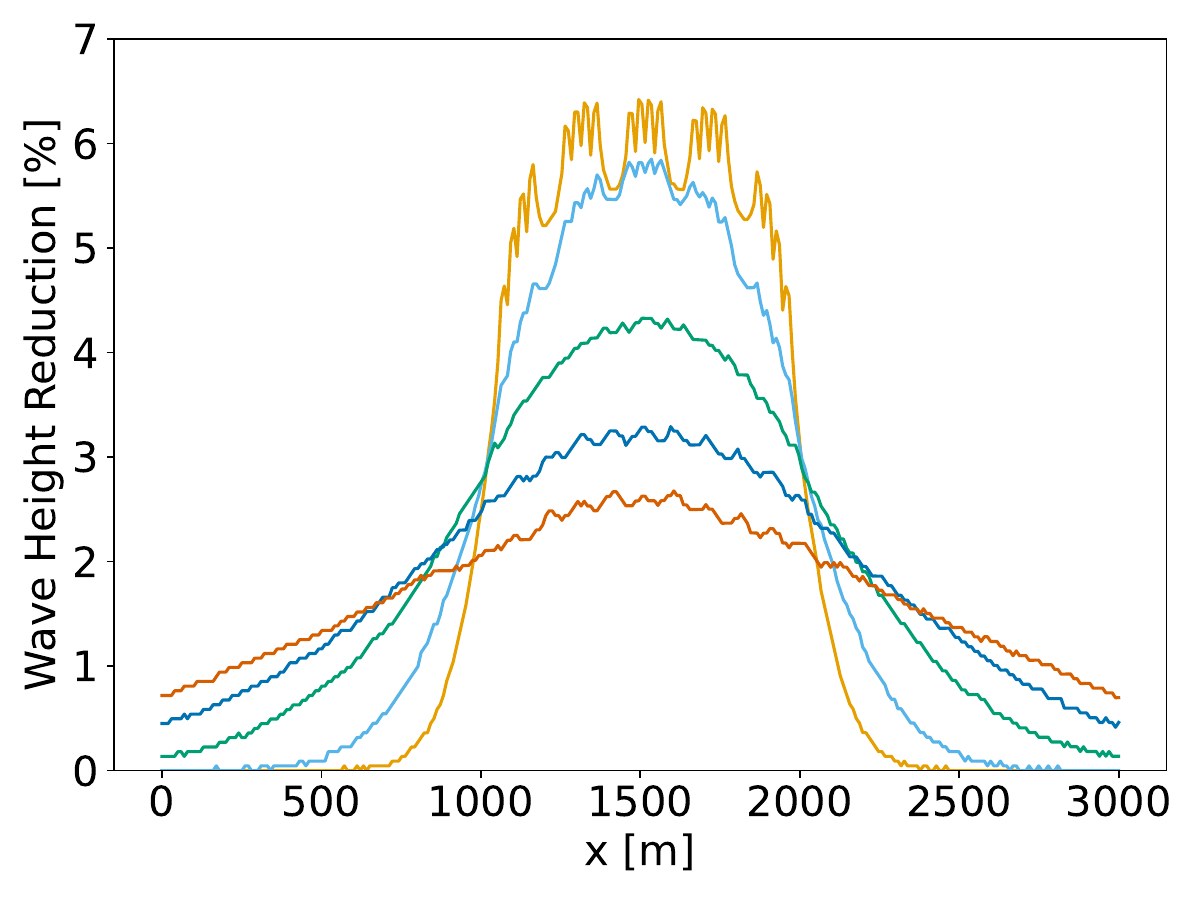}
         \caption{OSWEC array wave height reduction}
         \label{fig:osSWANwvht}
     \end{subfigure}
     \hfill
     \begin{subfigure}[b]{0.49\textwidth}
         \centering
         \includegraphics[width=\linewidth]{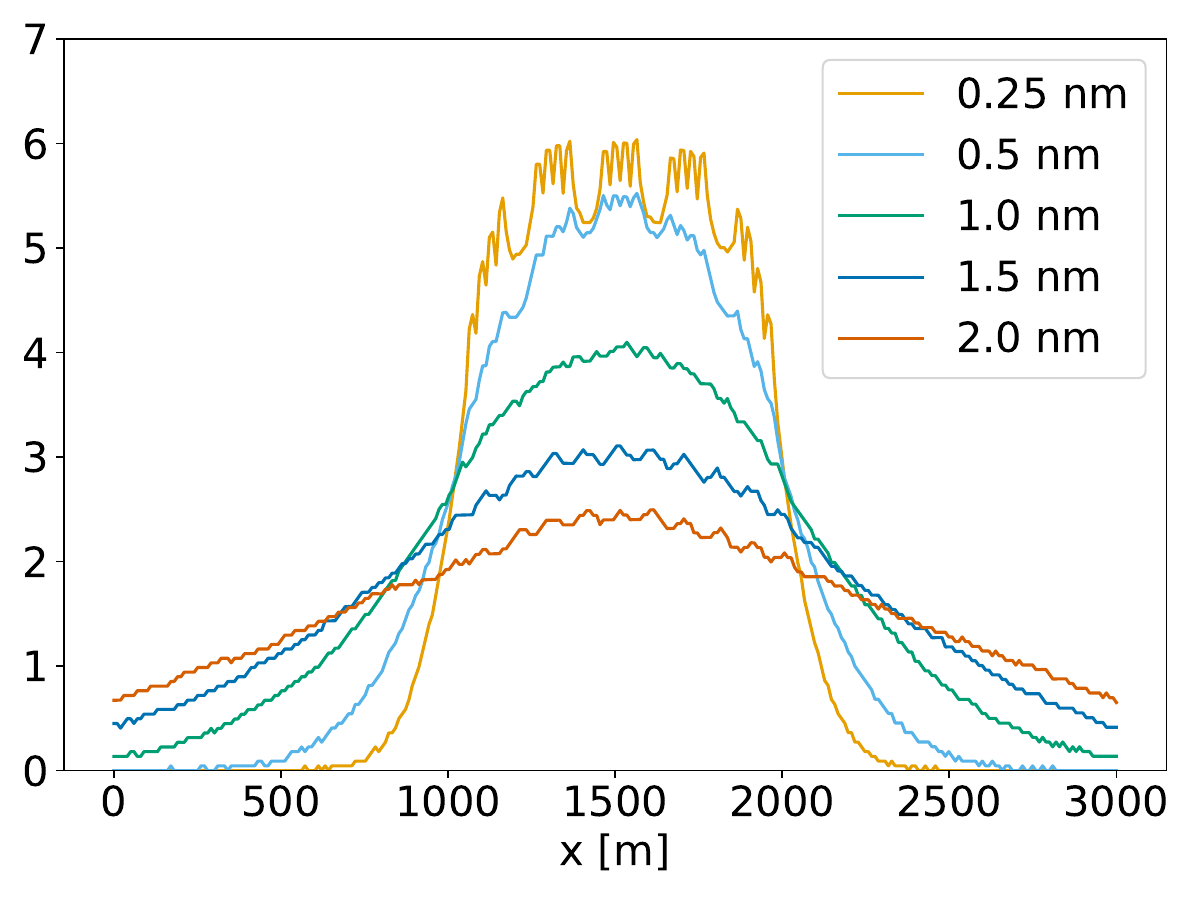}
         \caption{PA array wave height reduction}
         \label{fig:paSWANwvht}
     \end{subfigure}
    \caption{Wave height computed by the spectral balance solver in lee of a 20-body (a) OSWEC array and (b) PA WEC array. Red markers indicate WEC devices. (c) and (d) show the percent wave height reduction at five y-distances from the arrays.}
    \label{fig:shelter_SWAN}
\end{figure}

The wave sheltering performance was analyzed under oceanic conditions corresponding to the South Fork Wind Farm off the coast of Rhode Island. The farm consists of 12 turbines with a total farm rated power of 132~MW. A Pierson-Moskowitz spectra with a wind speed of 5.05~m/s at 19.5~m above sea level was used to represent the location. This corresponds to a significant wave height ($H_s$) of 2.19~m, a peak period ($T_p$) of 11~s, and an average period ($T_a$) of 5~s. The WECs were modeled in a 4-body farm to determine their individual sheltering capacities, including hydrodynamic interactions between devices. This 4-body unit was then repeated to produce a 20-body farm for the two different WEC architectures. These repeated units were placed 200~m apart, a large enough distance that interactions can be ignored\cite{BABARIT201368}. The wave height disturbance in lee of these farms over 4000~m is shown in Fig. \ref{fig:shelter_SWAN}. The disturbance is the ratio of the wave height when the array is present to the wave height without any obstructions, where a value less than one indicates a reduction in wave height.

Since the OSWECs' rated mechanical power extraction is nearly double that of the PAs, traditional power-capture modeling methods would predict the OSWEC to significantly outperform the PA in terms of wave sheltering. However, the PAs demonstrate only 0.4\% less wave height reduction with around a 50~m wider wake compared to the OSWECs. When accounting for array and fluid-structure interactions, the sheltering effect produced by two architectures is nearly identical. This has both policy and design implications for co-locating wind-wave systems. Currently, it is common practice to deploy an energy farm based on its total power production capacity. However, this does not accurately predict the WEC farm's sheltering capabilities. If sheltering is a primary driver of co-location, then cost, manufacturability, and co-location logistics should take precedent over power rating of the WEC device. 

\subsection*{Fatigue Damage Reduction}
\begin{figure}[ht!]
     \begin{subfigure}[b]{0.33\textwidth}
         \centering
         \includegraphics[width=\linewidth]{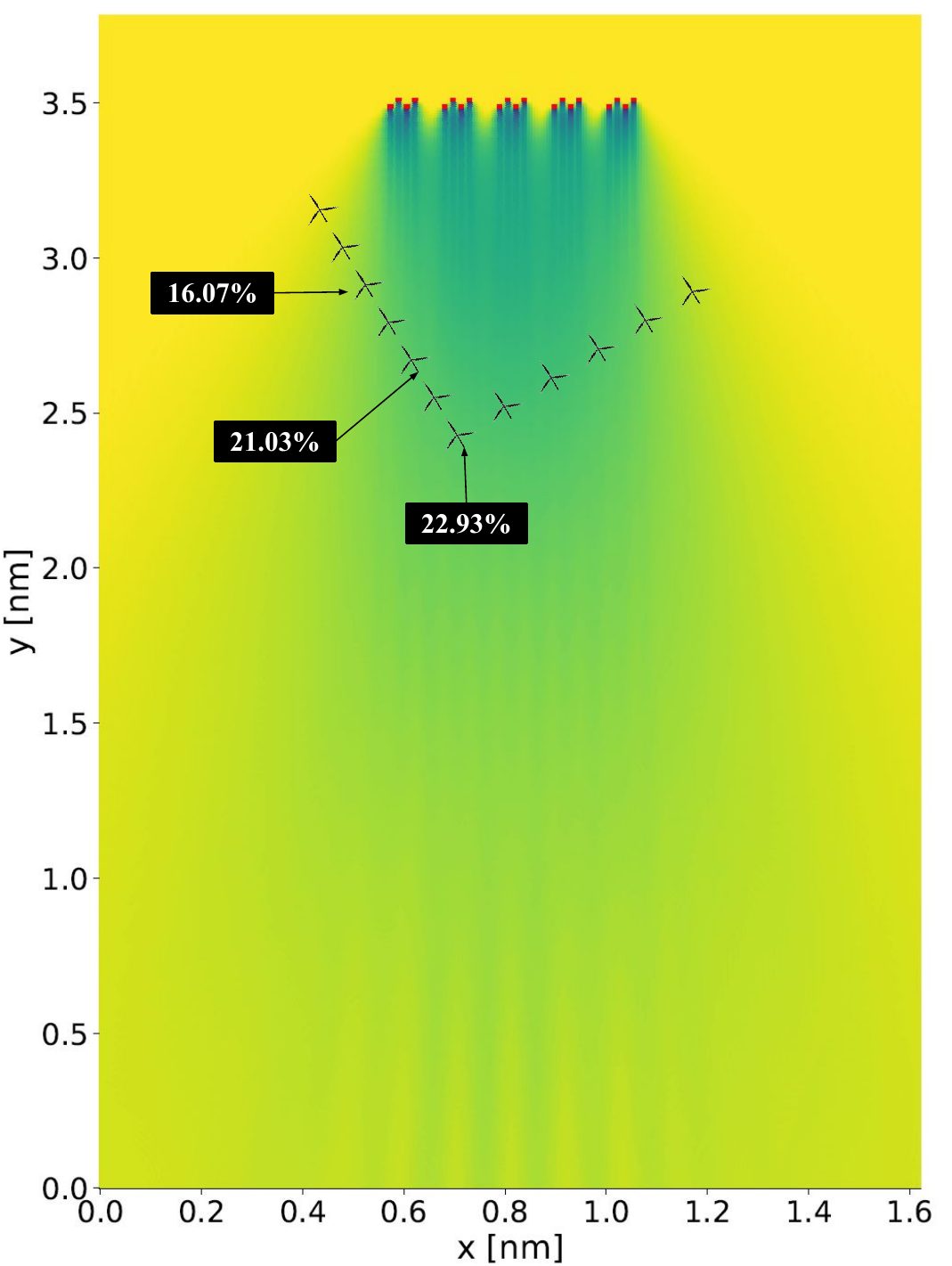}
         \caption{0.25 nautical miles}
         \label{fig:025nm}
     \end{subfigure}
     \hfill
     \begin{subfigure}[b]{0.29\textwidth}
         \centering
         \includegraphics[width=\linewidth]{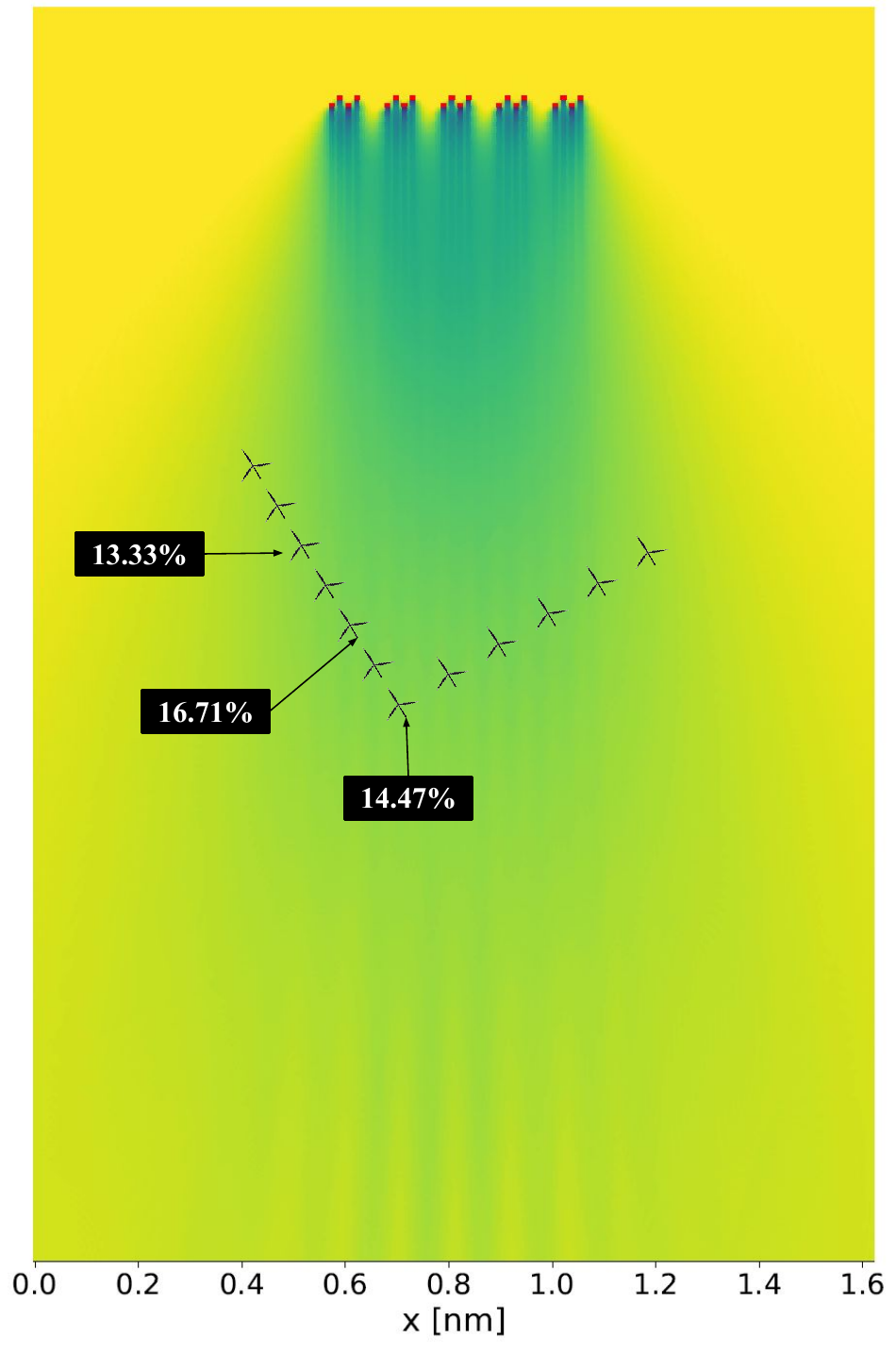}
         \caption{1.0 nautical miles}
         \label{fig:1nm}
     \end{subfigure}
     \hfill
     \begin{subfigure}[b]{0.34\textwidth}
         \centering
         \includegraphics[width=\linewidth]{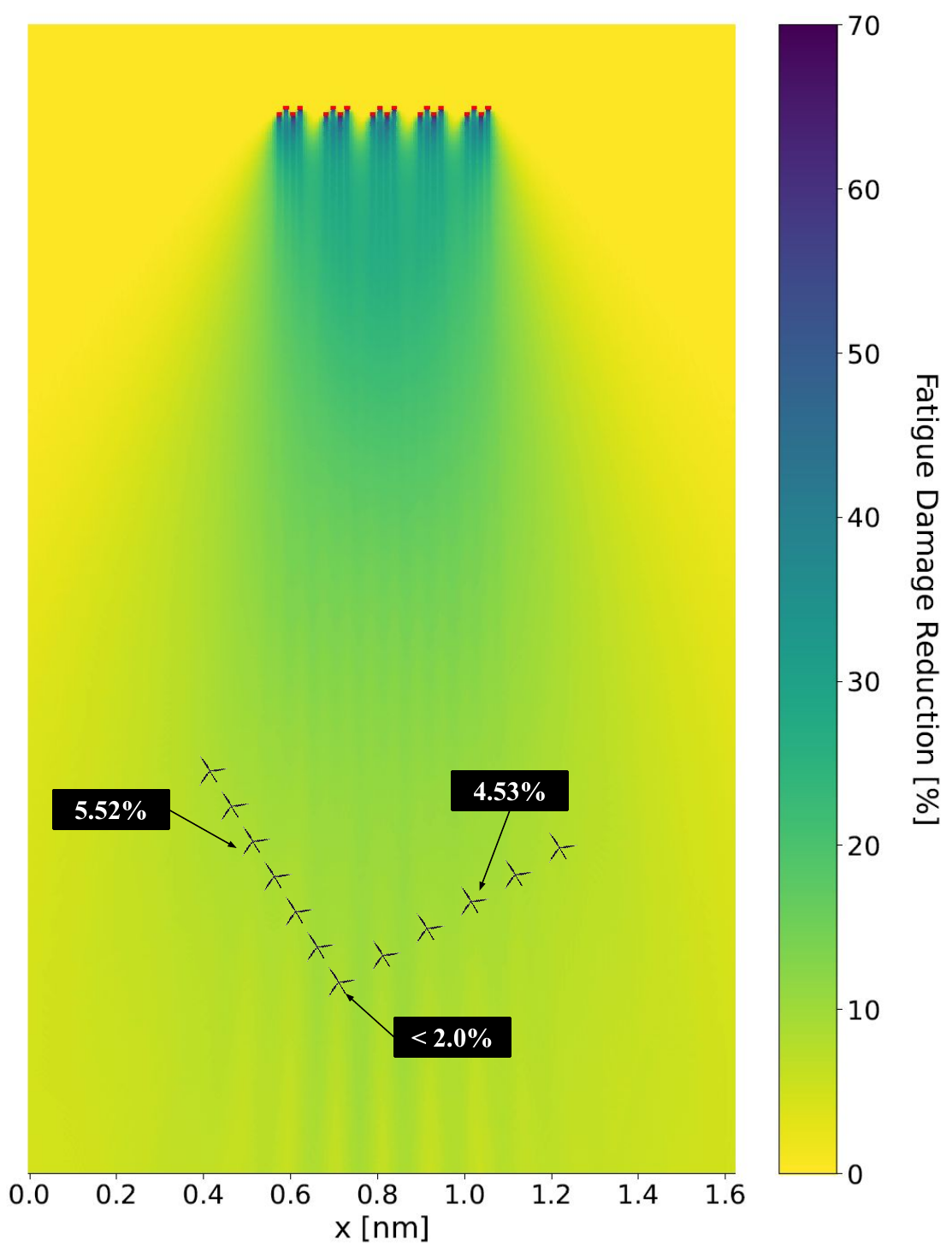}
         \caption{2.0 nautical miles}
         \label{fig:2nm}
     \end{subfigure}
    \caption{Depiction of wave-induced fatigue damage reduction at representative turbines in the South Fork wind farm due to the WEC farm. (a) shows results for when the start of the wind farm is 0.25~nm away from the WEC farm, (b) shows a distance of 1.0~nm, and (c) shows a distance of 2.0~nm.}
    \label{fig:turbines}
\end{figure}

Even small reductions in wave height provide large reductions in fatigue damage over the lifetime of an offshore wind turbine. The power in a wave ($P_{wave}$) is proportional to wave height squared ($H^2$), so a 5\% increase in wave height can result in up to 25\% greater power in the incident wave. This has major implications for wave-induced fatigue damage, in that a 5\% \textit{reduction} is wave height propagates to a 25\% reduction in fatigue damage over the turbine's lifetime. The 20-body farms achieved up to 6.42\% wave height reduction at 0.25~nm (463~m) in lee of the devices and between 1.0-2.68\% reduction at 2~nm (3,704~m). This corresponds to between 5.03-28.85\% reduction in fatigue damage on the turbine monopiles. Three potential co-location configurations of the South Fork wind farm and a WEC farm are shown in Fig. \ref{fig:turbines}. Placing the farms at 0.25~nm apart results in the largest fatigue damage reduction, but a reduced breadth of sheltering. This can be balanced by either adding more WECs to the farm to increase the sheltering width or by moving the turbines further downstream. Due to the South Fork wind farm configuration, the nearest turbine does not always experience the greatest sheltering. This points to a co-location optimization problem, designing the wind-wave farm to maximize the sheltering effect. The achievable reduction in fatigue could enable downsized monopiles to be feasible, resulting in between 20-60\% reduction in steel costs. Extending monopile lifetimes and reducing material requirements can reduce waste as well as the total cost of energy.

\section*{Discussion}
Hydrodynamic interactions were found to be relevant in determining device dynamics. Accounting for them gives a more accurate representation of expected device performance in the near- and far-field. Additionally, including fluid-structure interactions in the model exposes the bias present in traditional power-capture modeling methods for wave sheltering. Though the OSWECs are rated for a much higher power extraction level, they do not significantly outperform the PAs when sheltering waves. This suggests a potential tradeoff for hybrid wind-wave farm developers. The larger OSWECs may be more suitable for power smoothing in the hybrid farm due to their higher rated power, but the smaller PAs may provide the same amount of wave protection at a lower cost and simpler implementation.

Sheltering is greatest nearest to the array. This is an intuitive result, but the wave height reduction at larger distances should not be ignored. Even small reductions in wave height can significantly reduce wave-induced fatigue damage over an offshore wind turbine's lifetime, seeing a 10\% reduction in damage with only a 2\% reduction in wave height. This fatigue reduction can improve turbine lifetimes, or alternatively allow for smaller monopile diameter requirements. Reducing the monopile diameter by 1~m can result in a 20\% reduction in steel costs for the monopile, not including savings on manufacturing, transport, and installation. However, this is not distributed evenly at each monopile in the wind farm. To achieve maximum fatigue reduction, a wind-wave farm should be designed concurrently, with the turbine and WEC placement in careful consideration.

\subsection*{Implications for Cost and Deployment}
The reduction of fatigue damage due to wave loads propagates to reduced costs in capital costs (CAPEX) and operations and maintenance costs (OPEX), while the lifetime extension can reduces the levelized cost of energy (LCOE) of the offshore wind farm system. The National Renewable Energy Laboratory (NREL) released their 2024 Cost of Wind Energy Review, breaking down costs associated with deploying and operating onshore, offshore fixed-bottom, and offshore floating wind turbines\cite{NRELOWTcosts}. For offshore fixed-bottom, 16\% of the turbine costs are associated with materials for and the manufacturing of the monopile, making it the second largest cost driver of the overall turbine. Exportation and installation of the turbines accounts for 19.5\% of the total CAPEX of the farm, so reducing the size (and thereby mass) of the monopile propagates to reductions in these costs. Additionally, OPEX costs account for 26.9\% of total farm costs. Reducing damage on the turbines can reduce the amount of maintenance required for the farm, targeting a quarter of total costs. Finally, the LCOE is highly sensitive to the lifetime of the turbines, estimated to range between 15 and 35 years, with nominal being 25 years. Increasing the turbine lifetime from 25 to 35 years would produce an 11\% reduction in the cost of energy. Reducing the amount of required material and increasing the operational lifetime of these offshore systems targets large portions of the total costs of the system, up to 50\%. 

The co-location of offshore wind and WEC farms presents a unique synergistic opportunity. In addition to providing power smoothing for the intermittent wind resource and increasing operation and maintenance windows, they offer a lifetime extension and material cost reduction for offshore wind. This becomes increasingly relevant as wind moves further offshore, requiring protection from stronger waves. This coupling is also beneficial for the budding wave energy industry. Co-location offers the opportunity to share siting costs, mooring and transmission lines, and combine operation and maintenance teams. The results of this study indicate that co-located wind-wave farms present a promising pathway for the simultaneous achievement of fatigue load reduction and increased renewable energy generation. This synergy facilitates not only maintenance and cost savings, but positions offshore farms as adaptable, resilient energy hubs central to future distributed ocean power systems.

\section*{Methods}\label{methods}
\begin{figure}[t]
    \begin{subfigure}[b]{0.50\textwidth}
        \centering
        \includegraphics[width=\linewidth]{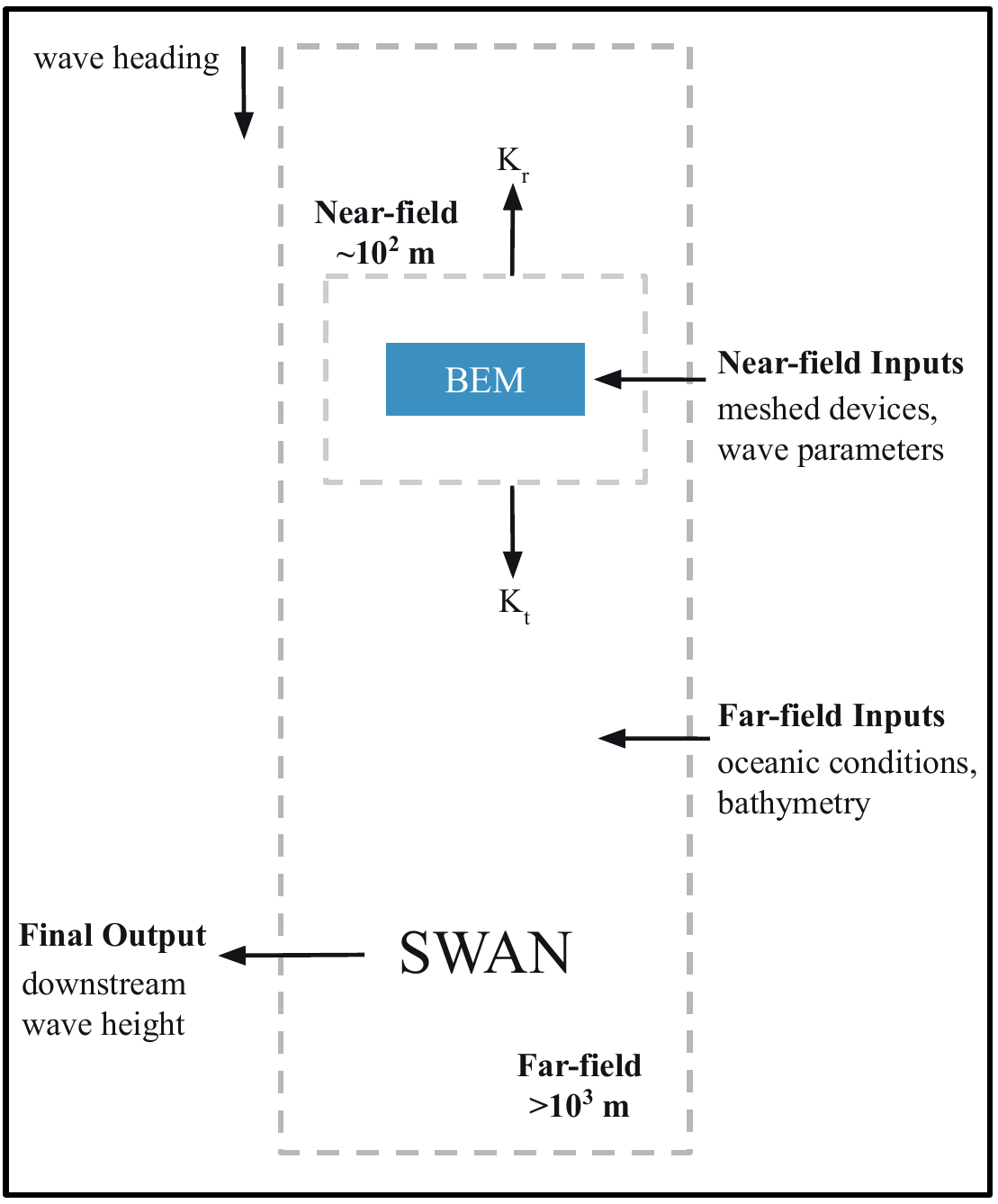}
        \caption{}
        \label{fig:numsys}
    \end{subfigure}
    \hfill
    \begin{subfigure}[b]{0.46\textwidth}
    \centering
        \includegraphics[width=\linewidth]{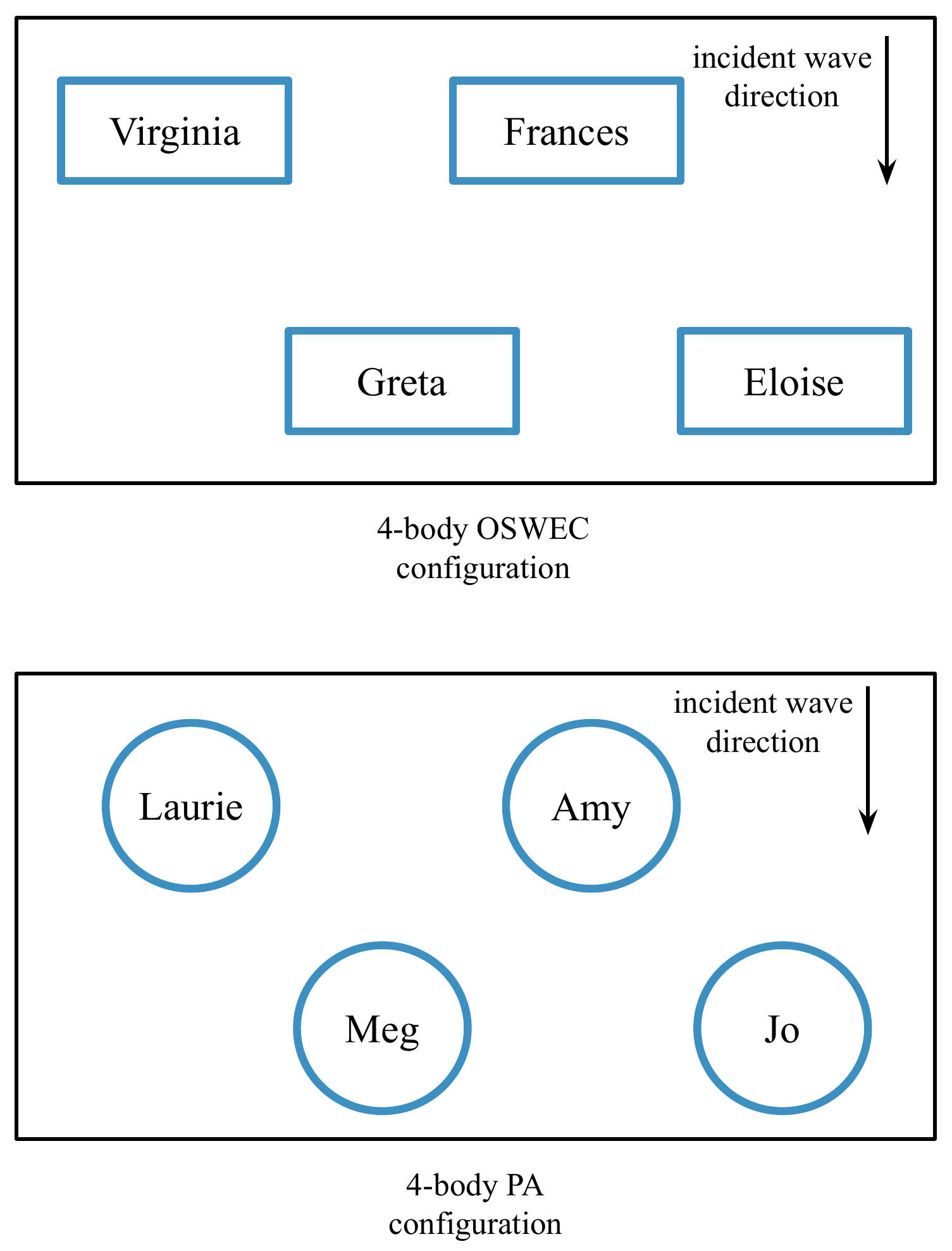}
        \caption{}
        \label{fig:arrayconfigs}
    \end{subfigure}
    \caption{(a) Graphical representation of hydrodynamic (near-field) BEM model interacting with the far-field spectral action model. (b) Four-body array configurations for oscillating surge WECs (top) and heaving point absorbers (bottom) that are meshed in the BEM solver.}
    \label{fig:numsys_arrayconfig}
\end{figure}

This study required identification of the following key metrics: each device's dynamic response (response amplitude operator, $RAO$), the transmission ($K_t$) and reflection ($K_r$) coefficients, the far-field wave height reduction, and the wind turbine fatigue damage. To quantify these metrics, an integrated model consisting of a near-field hydrodynamics solver and far-field wave propagation solver is required. The near-field solver computes the hydrodynamic properties of the WEC devices and the local disruption of the wave field. The $RAO$, $K_t$, and $K_r$ are found through this solver. This hydrodynamic model was validated through an experimental campaign, with the open-source dataset and post-processing scripts available at the \href{https://mhkdr.openei.org/submissions/637}{Marine and Hydrokinetic Data Repository}. This experimental validation bolsters the fidelity of the model, adding confidence to the results. The far-field solver computes the wave propagation over several thousand kilometers due to the presence of the array. It relies on $K_t$ and $K_r$ to define the energy reduction due to the array. The far-field wave height reduction is found through this solver. A visual depiction of how these solvers interact is shown in Fig. \ref{fig:numsys}. The wind turbine fatigue damage is found through Morrison wave force loading and the Palmgren-Miner rule. All software used to develop this work is available open-source, and the code used to generate the results can be found at the \href{https://github.com/symbiotic-engineering/transmission-reflection/releases/tag/hybrid_wind_wave}{SEA Lab Github Repository}. Further details on the model are provided in the following subsections.

\subsection*{Hydrodynamic Modeling}
WECs are modeled as second-order systems with some mass/inertia ($M_{ij}$ [kg and kg-m$^2$]), added mass due to the accelerating fluid ($A_{ij}$ [kg and kg-m$^2$]), radiation damping due to the device's forcing on the fluid ($B_{ij}$ [kg/s and kg-m$^2$/s]), and hydrostatic stiffness due to buoyancy ($K_{ij}$, [kg/s$^2$ and kg-m$^2$/s$^2$]). The are used to compute $\hat{X}_i$ [m], the complex body motion, shown as
\begin{equation}\label{RAO}
    \frac{\hat{X}_i}{A} = \frac{\hat{F}_{ex,i}}{-\omega^2(M_{ij} + A_{ij}) - j\omega B_{tot} + K_{ij}}
\end{equation}
where $A$ is the wave amplitude [m], $\omega$ is the wave frequency [rad/s], and $\hat{F}_{ex}$ is the wave exciting force on the body [N and N-m]. The "$i$" and "$j$" indices represent degrees of freedom of the body, ranging from 1 to 6. $B_{tot}$ represents the total damping in the system, including $B_{ij}$ and damping from the power take-off ($B_{PTO}$) based on Falnes' optimum control\cite{Falnes_Kurniawan_2020}, expressed as
\begin{equation}
    B_{PTO} = diag\left( \sqrt{B^2_{ij} + \left(\omega(M_{ij}+A_{ij}) - \frac{K_{ij}}{\omega}\right)^2} \right)
\end{equation}
These equations are for an isolated WEC floating body. However, in an array, the device's $\hat{X}_i$ equations are coupled, as their dynamic response depends on one another. To account for this in the model, each WEC device is allowed to oscillate in the single degree of freedom in which it extracts power. Then, the subscripts "$i$" and "$j$" in Eq. \ref{RAO} can be used to denote the WEC device number rather than degree of freedom. For example, $A_{12}$ is now the added mass in device $1$'s system due to device $2$'s oscillations. Accounting for all these couplings between devices is how the hydrodynamic interaction matrix is built and the coupled $\hat{X}_i$ of each device is determined. The response amplitude operator ($RAO$) is equivalent to the magnitude of the ratio between complex device motion and incident wave amplitude, shown by
\begin{equation}
    RAO_i = \left| \frac{\hat{X}_i}{A} \right|
\end{equation}
and the mechanical power ($P_{mech}$ [W]) is computed as\cite{Falnes_Kurniawan_2020}
\begin{equation}
    P_{mech} = \frac{1}{2}B_{PTO}|i\omega\hat{X}_i|^2
\end{equation}
The added mass and radiation damping matrices are determined through numerical methods. This study utilizes an open-source boundary element method (BEM) solver called Capytaine \cite{ancellin_capytaine_2019} to compute these hydrodynamic coefficient matrices. This method employs linear potential flow theory to solve the wave diffraction and radiation problems, assuming irrotational, incompressible, and inviscid flow. Neglecting viscous friction is a significant shortcoming of these models and is addressed further in the Experimental Validation section.

The BEM solver is also used to compute the near-field wave elevation [m]. The wave field is the superposition of the incident wave elevation ($\hat{\eta}_{in}$), the diffracted wave elevation ($\hat{\eta}_{diff}$), and the radiated wave elevation ($\hat{\eta}_{rad,i}$), shown as
\begin{equation}\label{wave_elevation}
    \hat{\eta} = \hat{\eta}_{in} + \hat{\eta}_{diff} + \sum_i \hat{\eta}_{rad,i} RAO_i
\end{equation}
The radiated wave elevation produced from every relevant degree of freedom must be included in this superposition. Additionally, the radiated waves must be multiplied by the device's $RAO$ to accurately represent their true magnitude. These computations are completed for individual frequencies (regular waves). However, the true ocean environment hosts multiple frequencies at once (irregular waves), with each wave frequency corresponding to a spectral density value. We represent this using a wave spectrum, and more details are provided in the Sheltering Characterization section.

\begin{figure}[ht]
     \begin{subfigure}[b]{0.45\textwidth}
         \centering
         \includegraphics[width=\linewidth]{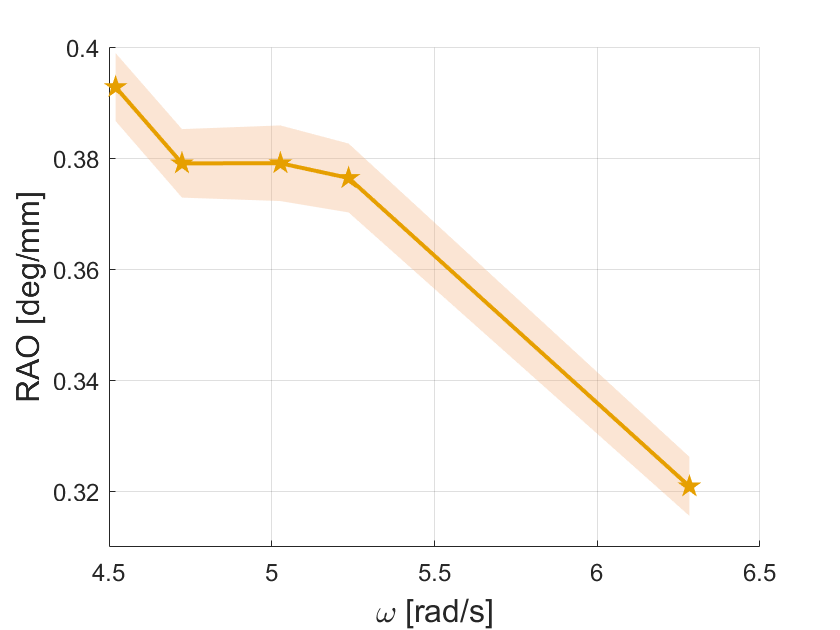}
         \caption{Isolated oscillating surge WEC}
         \label{fig:osiso}
     \end{subfigure}
     \hfill
     \begin{subfigure}[b]{0.45\textwidth}
         \centering
         \includegraphics[width=\linewidth]{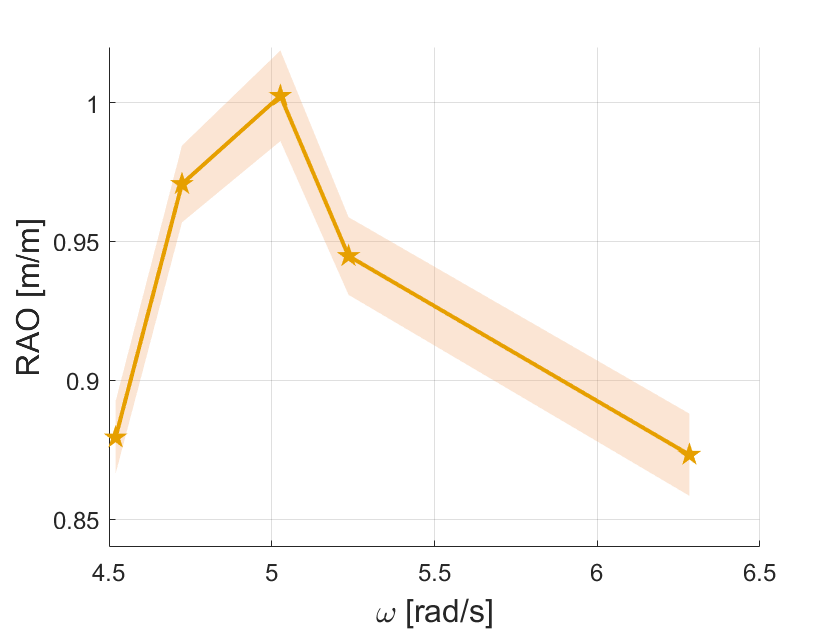}
         \caption{Isolated heaving point absorber WEC}
         \label{fig:paiso}
     \end{subfigure}
     \hfill
     \begin{subfigure}[b]{0.45\textwidth}
         \centering
         \includegraphics[width=\linewidth]{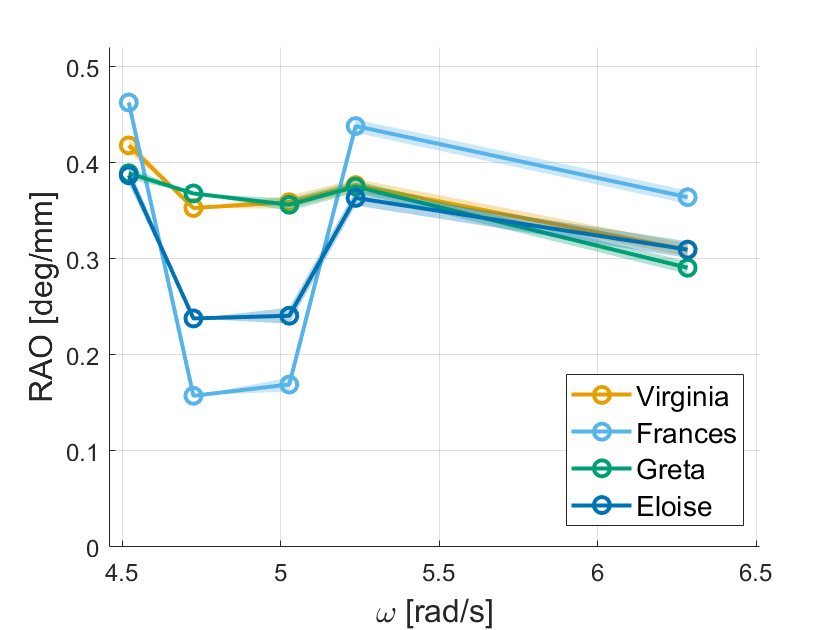}
         \caption{4-body oscillating surge WEC array}
         \label{fig: OSCWEC_RAO_valid}
     \end{subfigure}
     \hfill
     \begin{subfigure}[b]{0.45\textwidth}
         \centering
         \includegraphics[width=\linewidth]{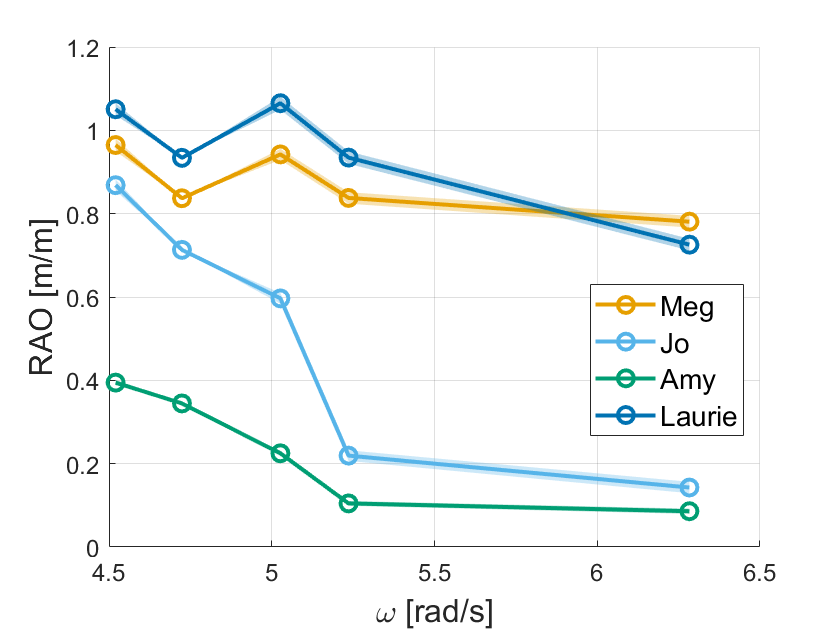}
         \caption{4-body point absorber WEC array}
         \label{fig:pahomrao}
     \end{subfigure}
    \caption{Device response amplitude operators for (a) an isolated heaving point absorber WEC, (b) an isolated oscillating surge WEC, (c) a 4-body point absorber WEC array, (d) a 4-body oscillating surge WEC array. The names in the legends correspond to physical devices from experimentation, shown in Fig. \ref{fig:arrayconfigs}. Shading around the lines indicates the 95\% confidence interval bounds on the data.}
    \label{fig:RAO_comparisons}
\end{figure}

Figure \ref{fig:arrayconfigs} depicts the two WEC array configurations analyzed in this report, one consisting of four PAs and the other of four OSWECs. These configurations were determined from a brute force Monte Carlo Method optimization of mechanical power production for fifteen different configurations over 12 sets of oceanic conditions. The configurations resembled common layouts found in literature as well as promising layouts from a previous optimization study\cite{khanalvitaledegoede}, ranging from 3- to 5-body arrays. The oceanic conditions corresponded to common peak periods found in offshore, deep-water locations (7 to 14~s) and the wave heading varied from 0 to 45\textdegree. The devices were also tested in isolation to quantify to what extent array interactions affected hydrodynamic performance. The array interactions were found to significantly alter each individual WEC's response to incident wave conditions. Figure \ref{fig:RAO_comparisons} shows the experimentally measured $RAOs$ of the isolated PA and OSWEC devices and the $RAOs$ of each device once they are operating in the array. The isolated PA device resembles a classic second-order system, with a peak response at $\omega$ = 5.03~rad/s. However, once operating in an array (Fig. \ref{fig:pahomrao}), a second peak was identified at $\omega$ = 4.72~rad/s. Additionally, the bandwidth of the response of each PA varied widely. A similar result is observed for the OSWECs, with a significant drop in $RAO$ between $\omega$ = 4.72 and 5.03~rad/s when placed in an array. This Froude 1:50, tank-scale experimental data was used to tune the BEM model. The BEM model at full ocean-scale ultimately produced the data in the presented analysis.

\subsection*{Sheltering Characterization}
\begin{figure}[ht]
    \centering
    \includegraphics[width=0.5\linewidth]{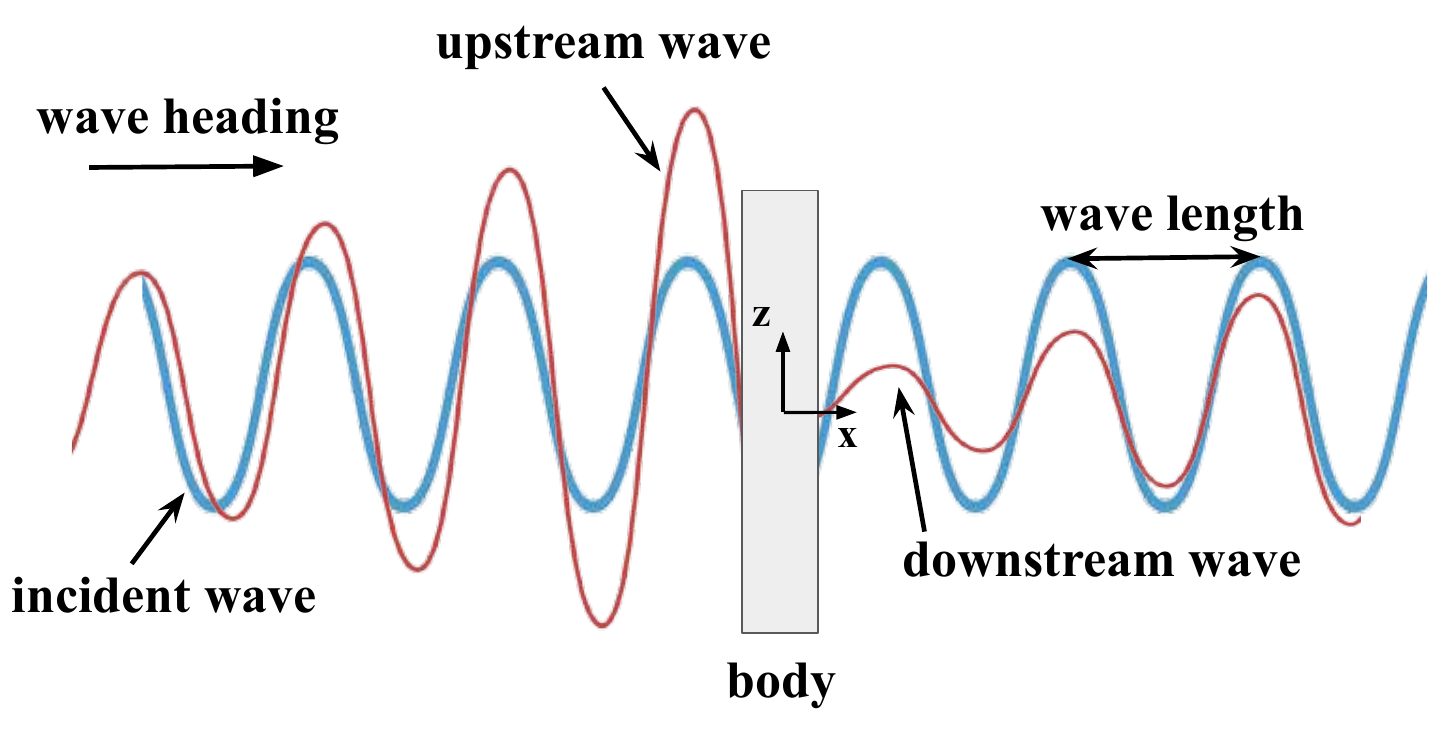}
    \caption{Representation of the incident, upstream, and downstream wave elevation with respect to the WEC device.}
    \label{fig:waveratios}
\end{figure}

To determine each device's effect on the wave field, the transmission ($K_t$) and reflection ($K_r$) coefficients were defined. The coefficients represent the amount of wave energy that transmits past the device and the energy that is reflected back upstream by the device, respectively. These coefficients are defined as
\begin{equation}
    K_t = \frac{|\hat{\eta}_{down}|}{|\hat{\eta}_{in}|}
\end{equation}
and
\begin{equation}
    K_r = \frac{|\hat{\eta}_{up}| - |\hat{\eta}_{in}|}{|\hat{\eta}_{in}|}
\end{equation}
where "up" and "down" denote the total wave elevation upstream and downstream of the device, respectively. A visual representation of these wave elevations with respect to the device is shown in Fig. \ref{fig:waveratios}. These coefficients were found numerically in the BEM solver. The average value of the wave elevation over one wavelength in front of and in lee of the device were used to determine the coefficients. The numerically computed free surface was validated with experimental data, with more details in the Experimental Validation section.

These coefficients were computed under regular wave inputs (i.e., one wave frequency at a time). To account for the spectral behavior (multiple wave frequencies at once) of the real ocean environment, the expected value of the coefficients in the wave spectra at the South Fork Wind Farm location was computed for each device. The Pierson-Moskowitz spectrum ($S(\omega)$ [$\frac{m^2}{rad/s}$]) was represented as\cite{HWANG2024118222}
\begin{equation}
    S(\omega) = H_s^2\frac{5}{16}\frac{\omega_p^4}{\omega^5}e^{-\frac{5}{4}(\frac{\omega_p}{\omega})^4}
\end{equation} 
where $\omega_p$ is the peak wave frequency [rad/s]. This spectrum is shown in the Appendix for reference. For each frequency, the significant wave height was multiplied by the corresponding transmission (or reflection) coefficient to represent the reduced (or increased) wave height due to a WEC device. This can be represented as
\begin{equation}
    \hat{S}_t(\omega) = S(\omega) K_t^2(\omega)
\end{equation}
where $\hat{S}(\omega)$ signifies the spectrum adjusted by either the transmission ($t$) or reflection ($r$) coefficients. The coefficient is squared because it is a multiplier to significant wave height ($H_s$), which is squared in the spectrum equation. The expected value for the altered significant wave height can be found using the zeroth moment ($m_0$) of the spectrum\cite{BOCCOTTI201543}:
\begin{equation}
    \hat{H_s} = 4\sqrt{m_0}
\end{equation}
where\cite{GHERBI2020505}
\begin{equation}
    m_0 = \int^\infty_{-\infty} \hat{S}(\omega) d\omega
\end{equation} Finally, the expected wave height reduction (or reflection) in irregular waves can be found by 
\begin{equation}
    K_{irr} = \frac{\hat{H_s}}{H_s}
\end{equation}
This form of the coefficients was used in the far-field model.

\subsection*{Far-Field Model}
Solving for the hydrodynamic properties of the devices helps describe the energy propagation through the system, but this method does not account for factors such as varying bathymetry, wind inputs, and other physical parameters of the ocean environment. Additionally, the BEM solver is only computationally efficient at resolving the free surface on the order of a few hundred meters\cite{ancellin}. WEC farms and offshore wind farms may be as far as several kilometers apart, requiring a different approach for determining the far-field effects on the wave environment. Therefore, a spectral approach is taken to model the far-field wave height. The spectral approach relies on statistical properties of the sea surface and linear wave theory. This modeling is done through the open-source software Simulating WAves Nearshore (SWAN) \cite{SWAN_sourceforge}. 

In SWAN, the devices are modeled as "obstacles" with a specified position, length, and transmission and reflection coefficient. The transmission ($K_t$) and reflection ($K_r$) coefficients defined above act as multipliers to the energy density at the grid lines that intersect with an obstacle line. Specific details on the numerical implementation of the coefficients can be found in the SWAN documentation \cite{SWAN_sourceforge}. A depiction of how the hydrodynamic and far-field solvers interact is shown in Fig. \ref{fig:numsys}.

The oceanic conditions selected for this work are based on the South Fork Wind Farm development, located about 19 miles southeast of Block Island, Rhode Island\cite{BOEM_SFlease}, spanning about 24~km$^2$. Bathymetry, significant wave height, and peak period data are used for the far-field wave reduction analysis. Bathymetry data was obtained through the Northeast Ocean data explorer \cite{NortheastOceanDataExplorer}, and wave conditions were obtained from NOAA Buoy 44097 \cite{NDBC44097}. This farm hosts 12 monopile turbines rated at 6-12 MW each \cite{TethysSouthForkWind} with a total farm rated power of 132~MW. The maximum monopile diameters are 10.97 m \cite{Denes2020}, which is the largest diameter before upgrading to more expensive jacket structures.

\subsection*{Wave Fatigue Damage}
The force the wave exerts on the monopile ($F_{wave}$) is quantified using the Morison equation \cite{Morison1950TheFE}, shown as
\begin{equation}\label{morison}
    F_{wave} = \frac{\pi}{4} d^2 C_M \rho_w \dot{u} + \frac{1}{2} C_D d \rho_w u |u|
\end{equation}
where $\rho_{w}$ is the density of sea water, $d$ is the monopile diameter, and $C_D$ and $C_M$ are the dimensionless drag and inertial coefficients, respectively. The drag coefficient is set to one and the inertial coefficient is set to two \cite{AgerschouEtAl}. The bending moment ($M_b$) is calculated as
\begin{equation}
    M_b = L F_{wave}
\end{equation}
where $L$ is the draft of the monopile. The axial forces are neglected\cite{VELARDE20173}, and the sectional stresses are entirely represented by the bending stress
\begin{equation}
    \sigma = \frac{y M_b}{I}
\end{equation}
where $y$ is the distance to the centroidal axis and $I$ is the monopile moment of inertia about the y-axis (pitch). The monopile wall thickness was chosen to be 0.125 m based on typical monopile design ranges \cite{escalera2024rapid,BMT2013}, and assumed constant along the submerged depth. The stress range ($\Delta \sigma$) is used to find the number of cycles till failure ($N_f$) using Wh\"oler's equation \cite{YETER2015518}
\begin{equation}\label{eq:cyc_to_fail}
    N_f = \bar{a} \Delta \sigma^{-m},
\end{equation}
where $\log(\bar{a})$ is the intercept of the $\log(N)$ axis on the S-N curve and $m$ is the material parameter. For this study, values from the DNV-RP-C203 \cite{dnv_rp_c203} where used. A low stress range was assumed along with $m = 5$, giving an experimentally determined $\log\bar{a}$ value of 13.617. Fatigue damage is then calculated using the Palmgren-Miner rule \cite{MINER}
\begin{equation}\label{miner}
    D = \sum_{i=1}^{n} \frac{n_i}{N_{f,i}},
\end{equation}
where $n$ signifies each load case (loads from wind, loads from wave, loads from different sea states, etc.), $n_i$ is the number of cycles of that load case, and $N_{f,i}$ is the number of cycles to failure at the stress range of that load case. This study only evaluates the percent of total fatigue damage due to wave loading over an offshore wind turbine lifetime of 25 years.

The cost of steel was modeled based on the volume of material required for the monopile as a function of monopile diameter and thickness. The price of steel was taken as the average reported values from an analysis of stainless steel costs in the European Union, China, and the United States in 2015\cite{GIULIODORI201512}.

\subsection*{Experimental Validation}
\begin{figure}[ht!]
    \centering
    \includegraphics[width=0.65\linewidth]{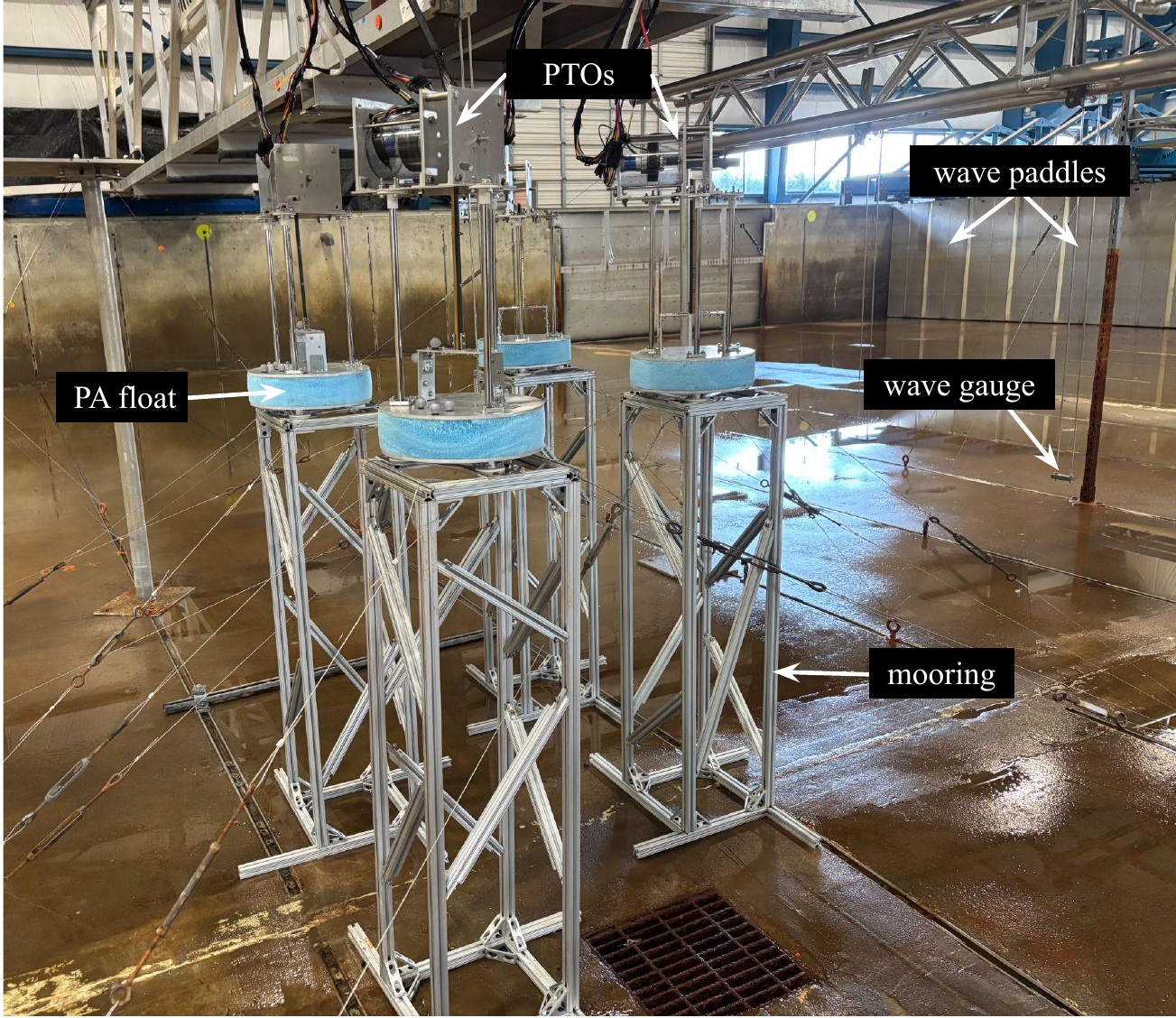}
    \caption{Four-body PA array setup in the O.H. Hinsdale Directional Wave Basin.}
    \label{fig:homoPAs}
\end{figure}

\begin{figure}[ht!]
     \begin{subfigure}[b]{0.40\textwidth}
         \centering
         \includegraphics[width=\linewidth]{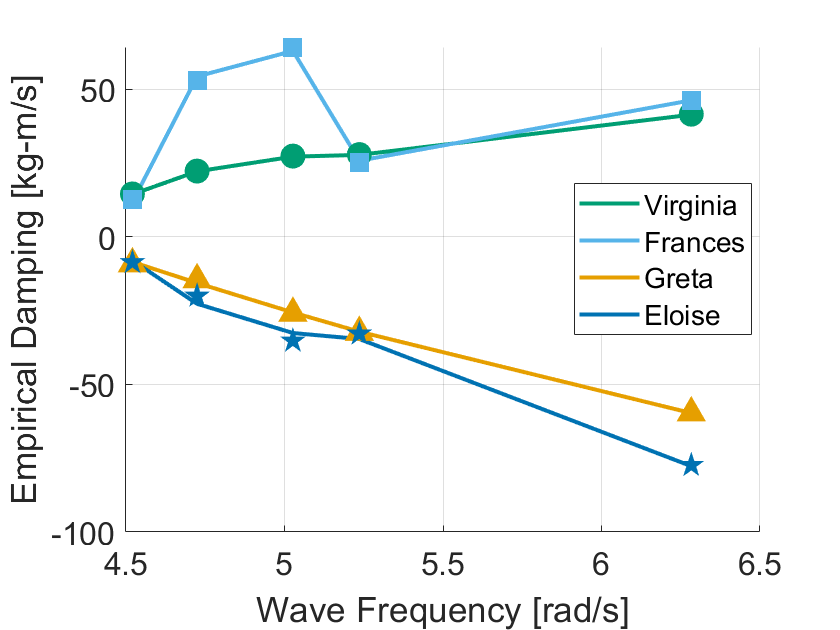}
         \caption{OSWEC empirical fit}
         \label{fig:OS_emp}
     \end{subfigure}
     \hfill
     \begin{subfigure}[b]{0.42\textwidth}
         \centering
         \includegraphics[width=\linewidth]{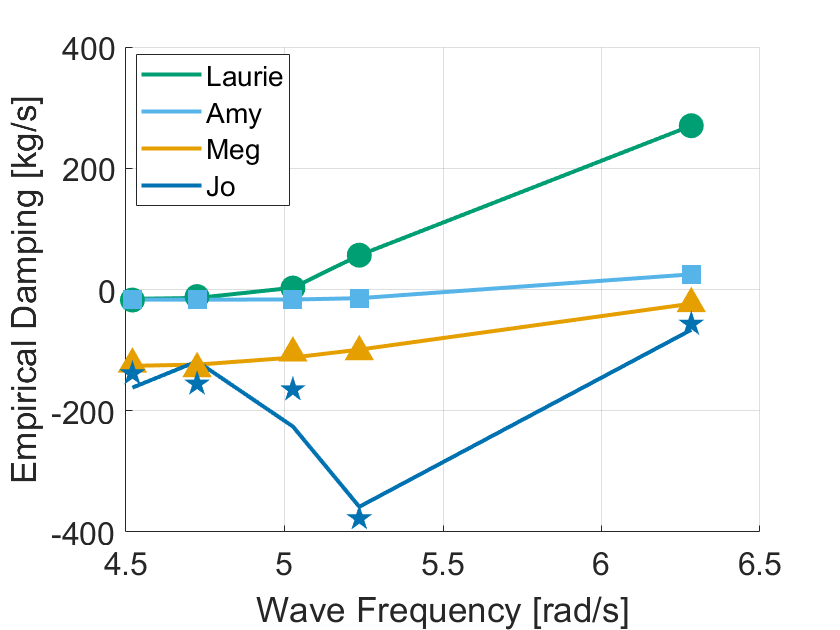}
         \caption{PA empirical fit}
         \label{fig:PA_emp}
     \end{subfigure}
     \begin{subfigure}[b]{0.42\textwidth}
         \centering
         \includegraphics[width=\linewidth]{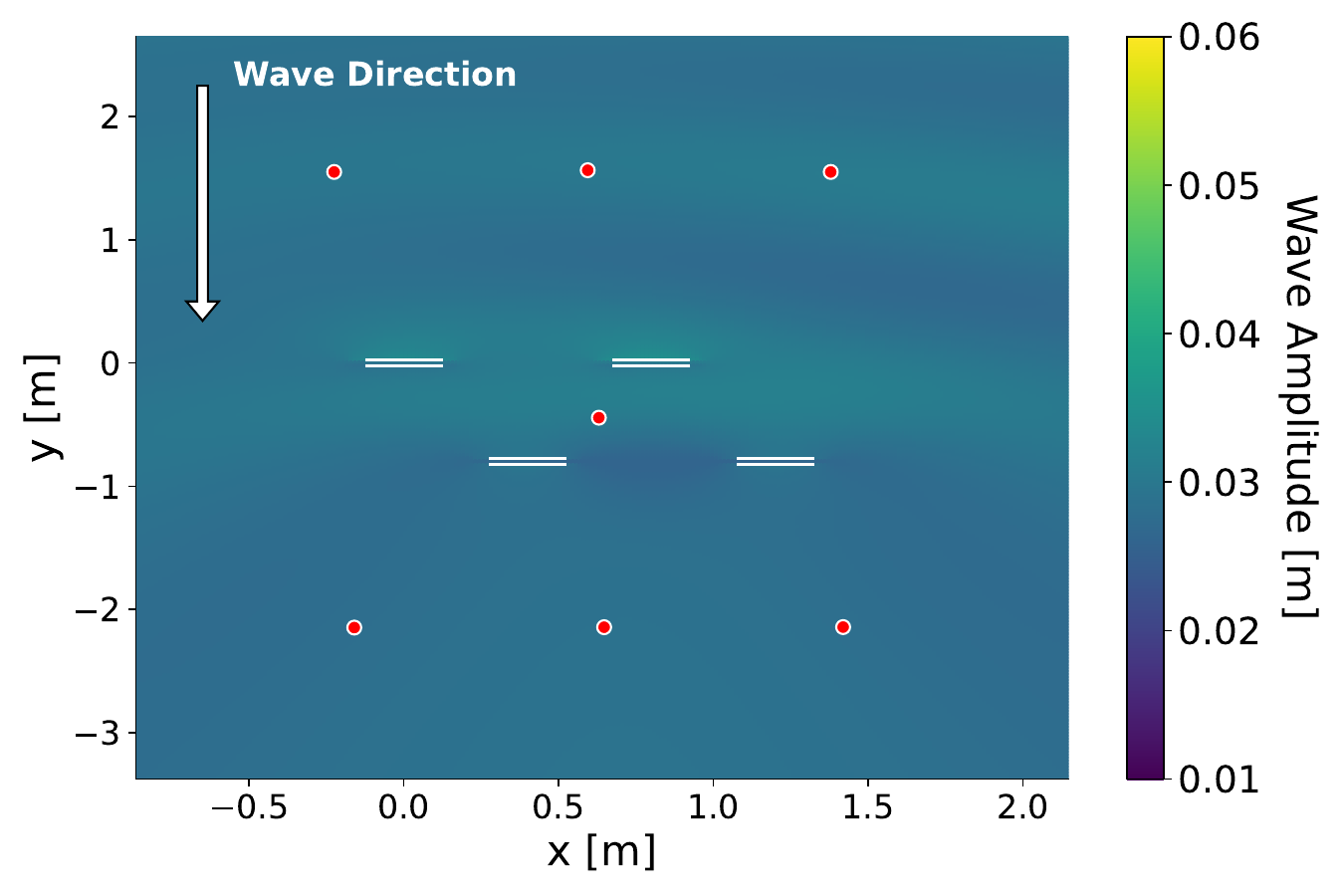}
         \caption{OSWEC wave field prediction}
         \label{fig:OSfield}
     \end{subfigure}
     \hfill
     \begin{subfigure}[b]{0.42\textwidth}
         \centering
         \includegraphics[width=\linewidth]{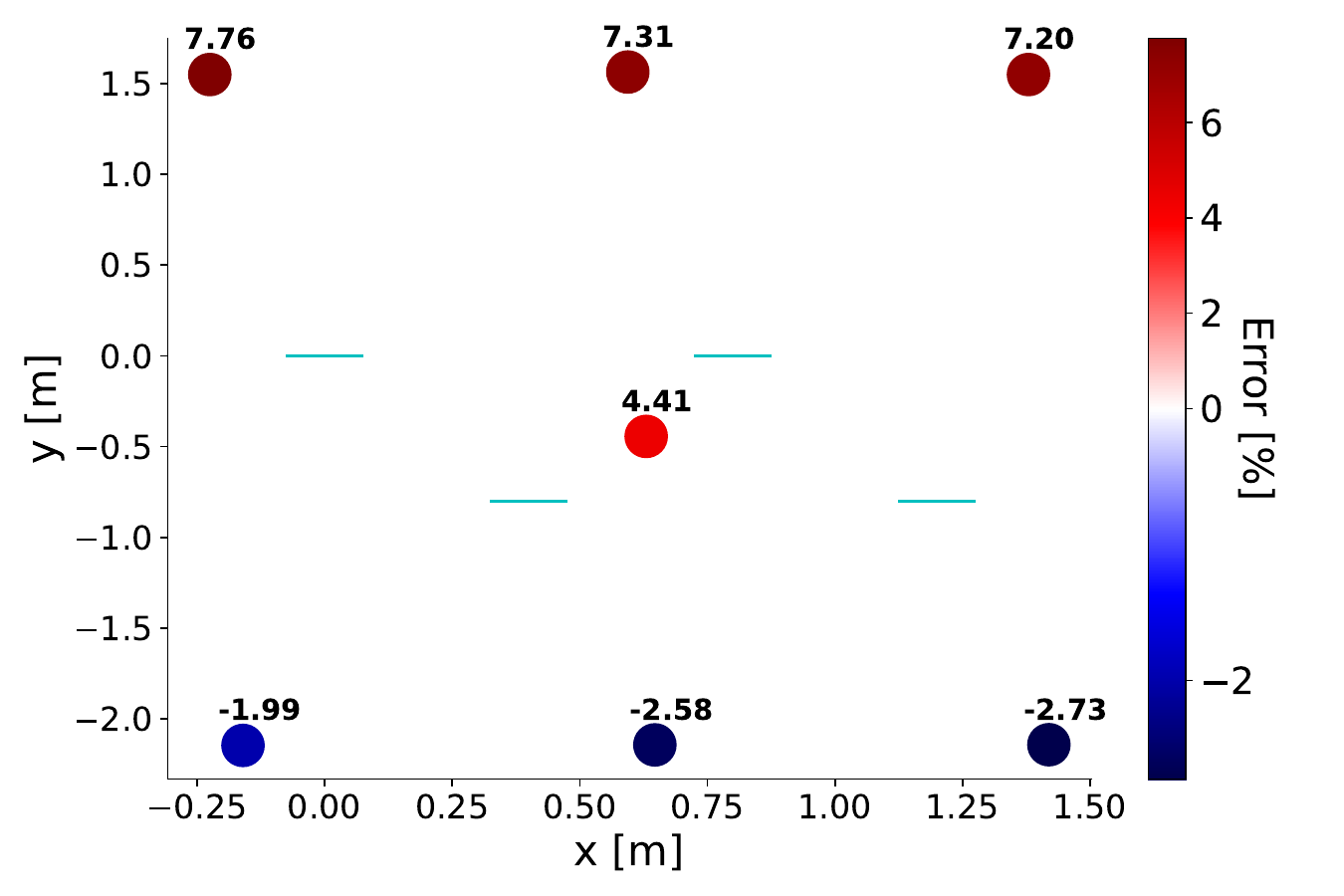}
         \caption{OSWEC wave field error}
         \label{fig:OSwaveerror}
     \end{subfigure}
     \hfill
     \begin{subfigure}[b]{0.42\textwidth}
         \centering
         \includegraphics[width=\linewidth]{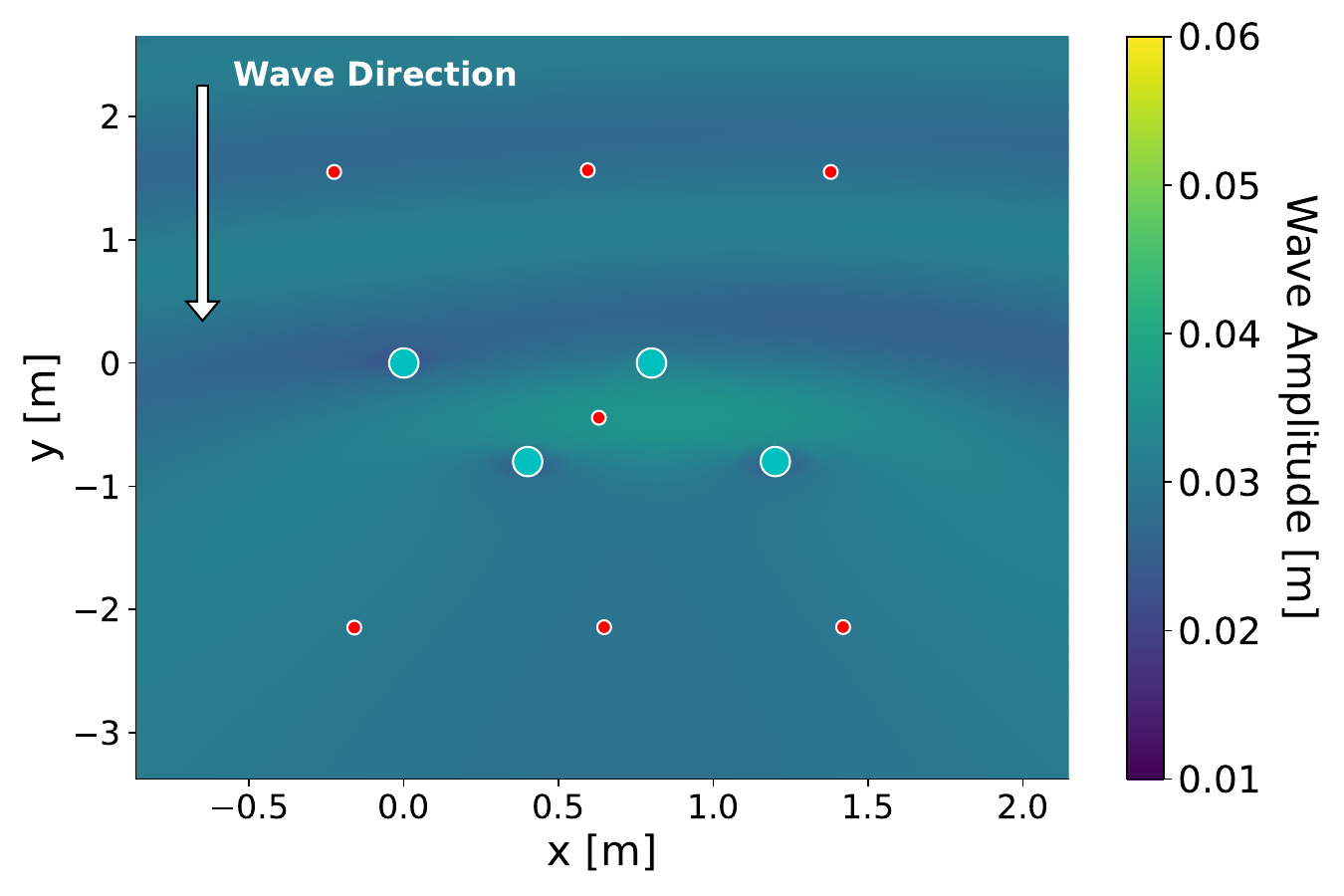}
         \caption{PA wave field prediction}
         \label{fig:PAfield}
     \end{subfigure}
     \hfill
     \begin{subfigure}[b]{0.42\textwidth}
         \centering
         \includegraphics[width=\linewidth]{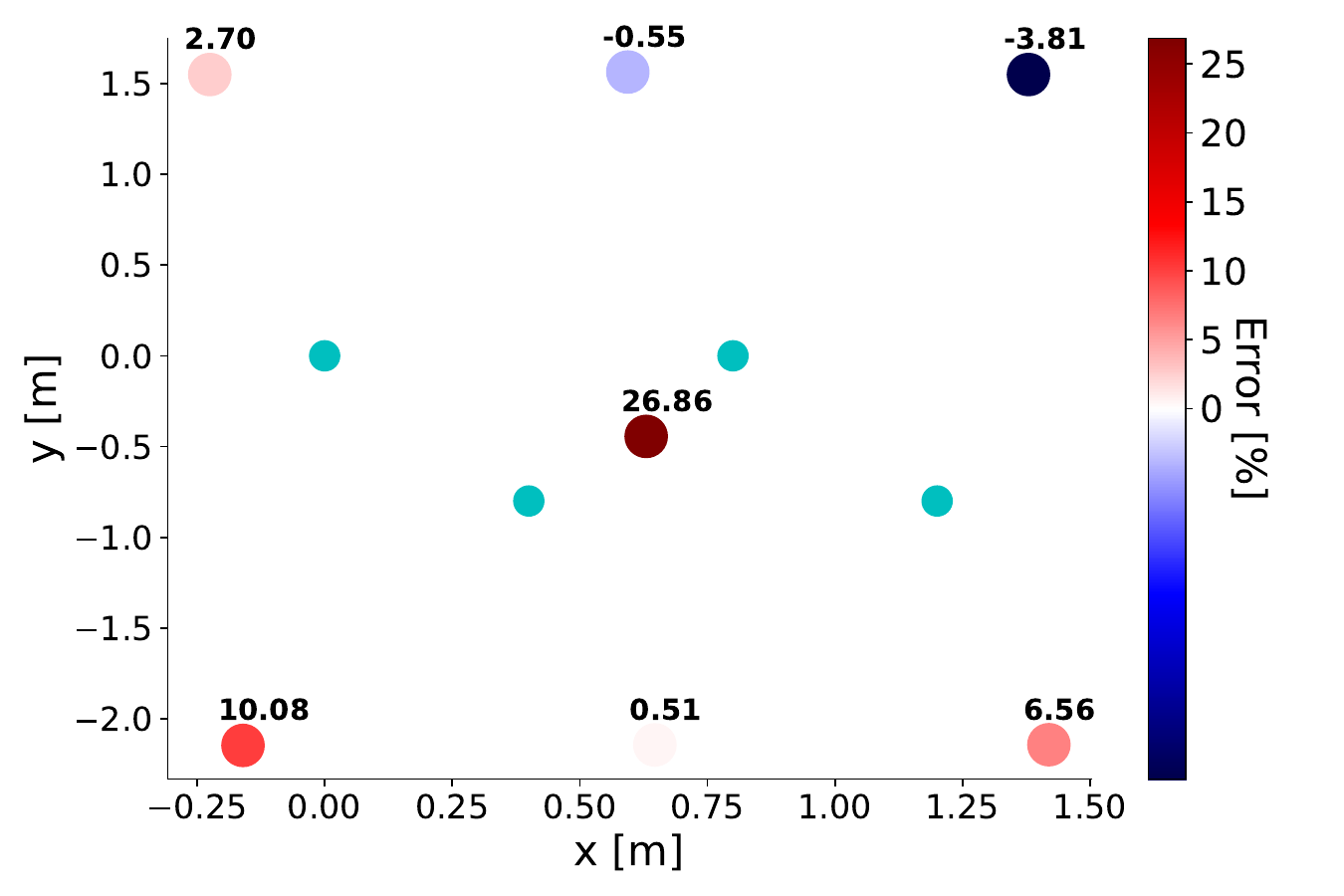}
         \caption{PA wave field error}
         \label{fig:PAwaveerror}
     \end{subfigure}
    \caption{(a) and (b) show curve fits for empirically derived damping terms for each device in the (a) OSWEC array and (b) PA WEC array. Solid lines indicate the curve fit while markers indicate the empirically derived coefficients. The curve fits have R$^2$ value greater than 0.9800, and the empirical data used to obtain the fits resulted in less than 1\% error propagation to the derived damping term. (c) and (e) show the wave amplitude computed by the BEM solver due to the presence of the WEC array (shown by light blue lines for OSWECs and circles for PAs). The red circles indicate locations of resistive wave gauges in the physical basin. (d) and (f) show the percent error at each probe. Values greater than zero indicate and over-prediction in wave amplitude, while values less than zero indicate an under-prediction.}
    \label{fig:exp_validation}
\end{figure}

Experimental validation was conducted using data from a deep water, non-breaking, non-Stokes fluid regime in the O.H. Hinsdale Directional Wave Basin. Experiments were conducted at the 1:50 scale through Froude scaling similarity laws. The PA array setup in the basin is provided in Fig. \ref{fig:homoPAs}.
To correct the numerical model's prediction of hydrodynamic response of the WEC systems, an empirical damping coefficient was introduced for each device in the array. This is an additional damping term, included in Eq. \ref{RAO} into the total damping term ($B_{tot}$). This empirical damping acts as a catch-all term for viscous effects that are not accounted for in the linear model as well as inaccuracies in the hydrodynamic coefficient computations. The results of this empirical fit are shown in Fig. \ref{fig:OS_emp} and \ref{fig:PA_emp}. Most fits were achieved with 1st, 2nd, or 3rd order polynomials with $R^2$ values greater than 0.9800. The empirical data used to obtain the fits resulted in less than 1\% error propagation to the derived damping term. Future work includes validation of individual hydrodynamic coefficients before fitting an empirical viscous damping term. 

The free surface directly around the devices (near-field) was validated using resistive wave gauges. The root-mean-squared (RMS) amplitude from the time-series data was compared to the BEM model free surface elevation computation. The results of this comparison are shown in Fig. \ref{fig:OSfield} through \ref{fig:PAwaveerror} for the scaled wave frequency of 4.52~rad/s. Only one frequency is shown for brevity, but results are discussed for frequencies between 4.52 to 6.28~rad/s, and their figures are shown in the Appendix. At full ocean-scale, this corresponds to wave periods between 7.07 and 9.83~s (0.889 and 0.639~rad/s). At the faster wave frequency, we observe slightly more error overall. This can potentially be attributed to the BEM assumption of small device characteristic dimension with respect to incident wavelength. Error is greatly improved at the slower wave period, similar to the dominant wave period analyzed in this study. However, we do observe over-estimation of the reflected wave height due to the OSWEC array. In terms of wave sheltering, the downstream wave height prediction is more important, as it dictates energy propagation towards the offshore wind turbines (or whatever structure is being sheltered). Therefore, we deem the downstream average error of 3.83\% for the OSWECs and 5.72\% for the PAs in the $\omega$ = 4.52~rad/s case to be acceptable.


\bibliography{references}

\section*{Data availability statement}
The open-source code can be found at the \href{https://github.com/symbiotic-engineering/transmission-reflection/releases/tag/hybrid_wind_wave}{SEA Lab Github Repository}, and the open-source experimental dataset can be accessed at the \href{https://mhkdr.openei.org/submissions/637}{Marine and Hydrokinetic Data Repository}.

\section*{Acknowledgments}
This work was funded in part by the Sea Grant Regional Research Project No.: R/ATD-18-NESG, the Cornell Atkinson Center for Sustainability through the 2023 Summer Mentored Research Program and the 2025 FAST Grant, and the U.S. Department of Energy through Testing and Expertise in Marine Energy (TEAMER) RFTS 12.

We would like thank Dr. Alaa Ahmed and Yashaswini Mandalam for their assistance in the experimental campaign and data processing.

\section*{Author contributions statement}
O.V. conceived the modeling techniques, executed model implementation, conducted experimental campaign, and developed data analysis techniques. O.V. and M.H. developed the numerical model, developed the experimental campaign, and analyzed results. O.V. and M.H. jointly acquired funding from the Cornell Atkinson Center for Sustainability and from the U.S. Department of Energy. M.H. acquired funding from the New York Sea Grant. All authors reviewed the manuscript.

\section*{Additional information}
Supplemental information can be found in the attached Appendix file. The authors declare no competing interests.

\end{document}